\documentclass{article}
\usepackage{arxiv}
\usepackage[T1]{fontenc}
\usepackage[utf8]{inputenc}
\usepackage{geometry}

\usepackage{latexsym}
\usepackage{graphicx}
\usepackage{amsfonts,amssymb,amsmath}
\usepackage{hyperref}

\usepackage{float}
\usepackage{cite}
\usepackage[nolist]{acronym} 
\usepackage{xcolor}
\usepackage{multirow}
\usepackage{placeins}

\usepackage{color}
\usepackage{subcaption}
\usepackage{hyperref}

\usepackage{tikz}
\usepackage{pgfplots}
\pgfplotsset{compat=1.16}
\usetikzlibrary{spy,calc}

\newif\ifblackandwhitecycle
\gdef\patternnumber{0}
\pgfkeys{/tikz/.cd,
    zoombox paths/.style={
        draw=orange,
        very thick
    },
    black and white/.is choice,
    black and white/.default=static,
    black and white/static/.style={ 
        draw=white,   
        zoombox paths/.append style={
            draw=white,
            postaction={
                draw=black,
                loosely dashed
            }
        }
    },
    black and white/static/.code={
        \gdef\patternnumber{1}
    },
    black and white/cycle/.code={
        \blackandwhitecycletrue
        \gdef\patternnumber{1}
    },
    black and white pattern/.is choice,
    black and white pattern/0/.style={},
    black and white pattern/1/.style={    
            draw=white,
            postaction={
                draw=black,
                dash pattern=on 2pt off 2pt
            }
    },
    black and white pattern/2/.style={    
            draw=white,
            postaction={
                draw=black,
                dash pattern=on 4pt off 4pt
            }
    },
    black and white pattern/3/.style={    
            draw=white,
            postaction={
                draw=black,
                dash pattern=on 4pt off 4pt on 1pt off 4pt
            }
    },
    black and white pattern/4/.style={    
            draw=white,
            postaction={
                draw=black,
                dash pattern=on 4pt off 2pt on 2 pt off 2pt on 2 pt off 2pt
            }
    },
    zoomboxarray inner gap/.initial=5pt,
    zoomboxarray columns/.initial=2,
    zoomboxarray rows/.initial=2,
    subfigurename/.initial={},
    figurename/.initial={zoombox},
    zoomboxarray/.style={
        execute at begin picture={
            \begin{scope}[
                spy using outlines={%
                    zoombox paths,
                    width=\imagewidth / \pgfkeysvalueof{/tikz/zoomboxarray columns} - (\pgfkeysvalueof{/tikz/zoomboxarray columns} - 1) / \pgfkeysvalueof{/tikz/zoomboxarray columns} * \pgfkeysvalueof{/tikz/zoomboxarray inner gap} -\pgflinewidth,
                    height=\imageheight / \pgfkeysvalueof{/tikz/zoomboxarray rows} - (\pgfkeysvalueof{/tikz/zoomboxarray rows} - 1) / \pgfkeysvalueof{/tikz/zoomboxarray rows} * \pgfkeysvalueof{/tikz/zoomboxarray inner gap}-\pgflinewidth,
                    magnification=3,
                    every spy on node/.style={
                        zoombox paths
                    },
                    every spy in node/.style={
                        zoombox paths
                    }
                }
            ]
        },
        execute at end picture={
            \end{scope}
            \node at (image.north) [anchor=north,inner sep=0pt] {\subcaptionbox{\label{\pgfkeysvalueof{/tikz/figurename}-image}}{\phantomimage}};
            \node at (zoomboxes container.north) [anchor=north,inner sep=0pt] {\subcaptionbox{\label{\pgfkeysvalueof{/tikz/figurename}-zoom}}{\phantomimage}};
     \gdef\patternnumber{0}
        },
        spymargin/.initial=0.5em,
        zoomboxes xshift/.initial=1,
        zoomboxes right/.code=\pgfkeys{/tikz/zoomboxes xshift=1},
        zoomboxes left/.code=\pgfkeys{/tikz/zoomboxes xshift=-1},
        zoomboxes yshift/.initial=0,
        zoomboxes above/.code={
            \pgfkeys{/tikz/zoomboxes yshift=1},
            \pgfkeys{/tikz/zoomboxes xshift=0}
        },
        zoomboxes below/.code={
            \pgfkeys{/tikz/zoomboxes yshift=-1},
            \pgfkeys{/tikz/zoomboxes xshift=0}
        },
        caption margin/.initial=4ex,
    },
    adjust caption spacing/.code={},
    image container/.style={
        inner sep=0pt,
        at=(image.north),
        anchor=north,
        adjust caption spacing
    },
    zoomboxes container/.style={
        inner sep=0pt,
        at=(image.north),
        anchor=north,
        name=zoomboxes container,
        xshift=\pgfkeysvalueof{/tikz/zoomboxes xshift}*(\imagewidth+\pgfkeysvalueof{/tikz/spymargin}),
        yshift=\pgfkeysvalueof{/tikz/zoomboxes yshift}*(\imageheight+\pgfkeysvalueof{/tikz/spymargin}+\pgfkeysvalueof{/tikz/caption margin}),
        adjust caption spacing
    },
    calculate dimensions/.code={
        \pgfpointdiff{\pgfpointanchor{image}{south west} }{\pgfpointanchor{image}{north east} }
        \pgfgetlastxy{\imagewidth}{\imageheight}
        \global\let\imagewidth=\imagewidth
        \global\let\imageheight=\imageheight
        \gdef\columncount{1}
        \gdef\rowcount{1}
        
    },
    image node/.style={
        inner sep=0pt,
        name=image,
        anchor=south west,
        append after command={
            [calculate dimensions]
            node [image container,subfigurename=\pgfkeysvalueof{/tikz/figurename}-image] {\phantomimage}
            node [zoomboxes container,subfigurename=\pgfkeysvalueof{/tikz/figurename}-zoom] {\phantomimage}
        }
    },
    color code/.style={
        zoombox paths/.append style={draw=#1}
    },
    connect zoomboxes/.style={
    spy connection path={\draw[draw=none,zoombox paths] (tikzspyonnode) -- (tikzspyinnode);}
    },
    help grid code/.code={
        \begin{scope}[
                x={(image.south east)},
                y={(image.north west)},
                font=\footnotesize,
                help lines,
                overlay
            ]
            \foreach \x in {0,1,...,9} { 
                \draw(\x/10,0) -- (\x/10,1);
                \node [anchor=north] at (\x/10,0) {0.\x};
            }
            \foreach \y in {0,1,...,9} {
                \draw(0,\y/10) -- (1,\y/10);                        \node [anchor=east] at (0,\y/10) {0.\y};
            }
        \end{scope}    
    },
    help grid/.style={
        append after command={
            [help grid code]
        }
    },
}

\newcommand\phantomimage{%
    \phantom{%
        \rule{\imagewidth}{\imageheight}%
    }%
}
\newcommand\zoombox[2][]{
    \begin{scope}[zoombox paths]
        \pgfmathsetmacro\xpos{
            (\columncount-1)*(\imagewidth / \pgfkeysvalueof{/tikz/zoomboxarray columns} + \pgfkeysvalueof{/tikz/zoomboxarray inner gap} / \pgfkeysvalueof{/tikz/zoomboxarray columns} ) + \pgflinewidth
        }
        \pgfmathsetmacro\ypos{
            (\rowcount-1)*( \imageheight / \pgfkeysvalueof{/tikz/zoomboxarray rows} + \pgfkeysvalueof{/tikz/zoomboxarray inner gap} / \pgfkeysvalueof{/tikz/zoomboxarray rows} ) + 0.5*\pgflinewidth
        }
        \edef\dospy{\noexpand\spy [
            #1,
            zoombox paths/.append style={
                black and white pattern=\patternnumber
            },
            every spy on node/.append style={#1},
            x=\imagewidth,
            y=\imageheight
        ] on (#2) in node [anchor=north west] at ($(zoomboxes container.north west)+(\xpos pt,-\ypos pt)$);}
        \dospy
        \pgfmathtruncatemacro\pgfmathresult{ifthenelse(\columncount==\pgfkeysvalueof{/tikz/zoomboxarray columns},\rowcount+1,\rowcount)}
        \global\let\rowcount=\pgfmathresult
        \pgfmathtruncatemacro\pgfmathresult{ifthenelse(\columncount==\pgfkeysvalueof{/tikz/zoomboxarray columns},1,\columncount+1)}
        \global\let\columncount=\pgfmathresult
        \ifblackandwhitecycle
            \pgfmathtruncatemacro{\newpatternnumber}{\patternnumber+1}
            \global\edef\patternnumber{\newpatternnumber}
        \fi
    \end{scope}
}

\usepackage{babel}
\usepackage{cite}
\usepackage{booktabs} 
\usepackage{xcolor} 
\usepackage{fancyvrb} 

\usepackage{booktabs} 

\date{\today}

\begin{document}

\title{Reverse image filtering using total derivative approximation and accelerated gradient descent}

\author{
 Fernando~J.~Galetto\thanks{Corresponding author.}\\
  Department of  Engineering \\
  La Trobe University\\
  Bundoora, VIC 3086, Australia\\
  \texttt{f.galetto@latrobe.edu.au} \\
     \And
  Guang~Deng \\
  Department of  Engineering \\
  La Trobe University\\
  Bundoora, VIC 3086, Australia\\
  \texttt{d.deng@latrobe.edu.au} \\
}
\begin{acronym}[AAAA ] 

\acro{IFM}{image formation model}
\acro{DCP}{dark channel prior}
\acro{IBM}{image blur map}
\acro{CNN}{convolutional Neural network}
\acro{GIF}{guided image filter}
\acro{DoF}{depth of field}
\acro{SDoF}{shallow depth of field}
\acro{CV}{computer vision}
\acro{LTI}{linear translation-invariant}
\acro{GF}{guided filter}
\acro{BF}{bilateral filter}
\acro{RGF}{rolling guidance filter}
\acro{WGIF}{weighted guided image filter}
\acro{NLT}{non linear transformation}
\acro{NN} {neural network}
\acro{CNNs}{convolutional neural networks}
\acro{PSF}{point spread function}
\acro{GS}{Gaussian filter} 
\acro{AMF}{adaptive manifolds filter}
\acro{LoG}{Laplace-of-Gaussian}
\acro{L0}{image smoothing via $L_0$ minimization}
\acro{RTV}{structure extraction from texture via relative total variation}
\acro{ILS}{image smoothing via iterative least squares}
\acro{KNN}{K-nearest neighbour}
\acro{UM}{unsharp masking}
\acro{ZRF}{zero-order reverse filtering}
\acro{FRF}{iterative first-order reverse image filtering}
\acro{AGD}{accelerated gradient descent}
\acro{GD}{gradient descent}
\end{acronym}

\maketitle

\begin{abstract}
In this paper, we address a new problem of reversing the effect of an image filter, which can be linear or nonlinear. The assumption is that the algorithm of the filter is unknown and the filter is available as a black box. We formulate this inverse problem as minimizing a local patch-based cost function and use total derivative to approximate the gradient which is used in gradient descent to solve the problem. We analyze factors affecting the convergence and quality of the output in the Fourier domain. We also study the application of accelerated gradient descent algorithms in three gradient-free reverse filters, including the one proposed in this paper. We present results from extensive experiments to evaluate the complexity and effectiveness of the proposed algorithm. Results demonstrate that the proposed algorithm outperforms the state-of-the-art in that (1) it is at the same level of complexity as that of the fastest reverse filter, but it can reverse a larger number of filters, and (2) it can reverse the same list of filters as that of the very complex reverse filter, but its complexity is much smaller.

\end{abstract}
\keywords{
Inverse filtering, optimization, accelerated gradient descent, total derivative.
}

\section{Introduction}

Solving inverse problems is of critical importance in many science and engineering disciplines. In image processing, a typical problem is called image restoration \cite{gonzalez2004digital} in which the distortion is modelled by a linear shift invariant (LSI) system and additive noise. Other well-known problems include:  dehazing\cite{he2010single,li2017single} and denoising \cite{Buades2005, dabov2007image, jain2016survey}. In these applications, an edge-aware filter such as \cite{tomasi1998bilateral, he2012guided, wei2018joint, yin2019side, Sun2020, li2015weighted} plays an important role. In general, solutions to such problems are model based. For example, in image restoration an observation model with a known or unknown blurring filter kernel is assumed, while in dehazing, a physical image formation model is also assumed. An essential part of such algorithms is to estimate the model parameter such as the filter kernel in image restoration or the transmission map in dehazing.  Recently, deep neural networks are used in solving inverse problem in imaging applications such as inverse tone mapping \cite{endoSA2017} and de-blurring \cite{CNN_deblur2021}, where a neural network is trained to solve a specific inverse problem. Other notable examples of solving inverse problems are computational imaging techniques, such computed tomography \cite{CT1988} and compressive sensing \cite{CompressedSensing}. But they are not related to our current research.

The increasing use of many image processing filters in many areas has led to a new formulation of the inverse problem. Let $g(.)$ represent an image filter which is usually available as a software tool. The user of the filter has no knowledge of the exact algorithm which it implements. As such, the filter can be regarded as an available black box which takes an input image $\boldsymbol{x}$ and produces an output image $\boldsymbol{b}=g(\boldsymbol{x})$. The new inverse problem is thus to estimate the original image given the observation $\boldsymbol{b}$ and the black box filter $g(.)$. Two factors make the study of such a problem theoretically and practically important.

\begin{itemize}

    \item Traditional methods, which rely on using gradient information to solve optimization problems, will face a challenge of lack of such information because the filter is unknown. New algorithms, which do not directly rely on gradient information, must be developed. 

    \item While classical image restoration techniques are mostly dealing with known or estimated linear filter kernels for the inversion, solutions of the new inverse problem, which do not rely on the model of the filter to be inverted, can potentially deal with a wide range of unknown linear and nonlinear filters. 

\end{itemize}

A brief review of current work on this new problem is presented in section \ref{sec:previous work} which includes three spatial domain methods: the T-method \cite{tao2017zero}, R-method  \cite{milanfar2018rendition} and P-method \cite{belyaev2021two} and a frequency domain method called F-method \cite{dong2019iterative}. Except the T-method, which can be motivated as a fixed-point algorithm, the other three share a similar idea of formulating a cost function and use gradient approximation to derive the solution to the optimization problem.  These algorithms have been used to reverse a wide range filters including some traditional filters \cite{gonzalez2004digital} such as: Gaussian filter, disk averaging filter, motion blur filter,  \ac{LoG}, and some edge-aware filters such as: \ac{RGF} \cite{zhang2014rolling},  \ac{AMF} \cite{gastal2012adaptive}, \ac{RTV} \cite{xu2012structure}, \ac{ILS} \cite{liu2020real}, \ac{L0} \cite{xu2011image}, \ac{BF} \cite{tomasi1998bilateral},  self \ac{GF}  \cite{he2012guided} and \ac{GF} with the guidance image generated from filtering an image by a Gaussian filter. 

This work is motivated by the P-method and R-method. Our goal is to develop a principle-based new algorithm which is of low complexity and is highly effective (able to reverse the effect of a wide range of filters). This is in contrast to a deep neural network approach in which a network is trained to deal with a specific problem. The basic idea for our development is similar to that of the P- and R-method. The main differences are in the cost function formulation and gradient approximation. While authors of the P- and R-method use a global cost function, we use a patch-based local cost function. We have also used total derivative approximation (TDA) of the gradient. This differs from the approximation approach of the P- and R-method.  The new algorithm is thus called the TDA-method. In addition, we have studied the application of well-known  accelerated gradient descent (AGD) methods \cite{optimizationAlgo} which are listed in Table \ref{tab:AGD_table_summary}. AGD methods are commonly used in machine learning \cite{Goodfellow-et-al-2016}, but their application have not been explored in this context. We have conducted extensive experiments to show that while the proposed TDA-method is of the same level of complexity as that of the T-method, it is as effective as the P-method but at a much lower complexity.  

Key contributions of this work are as follows. 

\begin{itemize}

    \item In section 3, we present the development of the TDA-method (section \ref{sec:TDA}) and provide a theoretical analysis of the convergent condition and factors related to the quality of the recovered image (section \ref{sec:convergence}). We then discuss applying AGD methods to 3 spatial domain reverse filtering algorithms, including the TDA-method (section \ref{sec:AGD}). 

    \item In section 4 we show results of extensive experimental study of the proposed method, including: the effect of parameter settings such as learning rate and number of iterations (section \ref{sec:para}), qualitative and quantitative evaluations and comparisons to demonstrate the performance of current works and the TDA-method in reversing a wide range of filters (section \ref{sec:applications}) and a study of using AGD methods (section \ref{sec:AGD results}). We also present an intriguing case study where all methods failed to reverse the effect of a simple median filter to illustrate the limitation of this class of gradient-free algorithms (section \ref{sec:limitation}). 

\end{itemize}

We adopt the following notations. An image is represented as a matrix $\boldsymbol{x}$ and the filter denoted $g(.)$ operates on the image and produces the observed image $\boldsymbol{b}=g(\boldsymbol{x})$. The subscript is used to represent iteration index and arithmetic operations are conducted pixel-wise.  $||\boldsymbol{x}_k||$ is the 2-norm of the image at $k$th iteration. The Fourier transform of $\boldsymbol{x_k}$ is represented as $\mathcal{F}(\boldsymbol{x}_k)$.

\section{Previous work}\label{sec:previous work}

We follow the terminology used in \cite{belyaev2021two} which called a particular reverse filter a method and briefly comment on each of them in this section. We present a summary of related previous work in Table \ref{tab:compareAlgo} where we define the following variables which will be used in rest of this paper. 

\begin{itemize}

    \item $\boldsymbol{q}_k = \boldsymbol{b} - g(\boldsymbol{x}_k)$.  

\item $\boldsymbol{p}_k = g(\boldsymbol{x}_{k}+\boldsymbol{q}_{k})-g(\boldsymbol{x}_{k}-\boldsymbol{q}_{k})$. 

\item $\boldsymbol{t}_k=g(\boldsymbol{x}_{k}+\boldsymbol{q}_{k})-g(\boldsymbol{x}_{k})$.

\end{itemize}

The T-method, which is originally called zero-order reverse filter \cite{tao2017zero}, can be formulated as solving a fixed point iteration problem \cite{ortega2000iterative}  $\boldsymbol{x}=\boldsymbol{x}+\boldsymbol{b}-g(\boldsymbol{x})$. It is the fastest method among all reverse filtering methods considered in this paper. When the contraction mapping condition ($|1-g’(x)|<1$ in 1D case) is satisfied for a particular filter, the T-method produces good results. Otherwise the iteration is unstable, leading to unbounded results. An unstable example is the reverse of an average filter using the T-method. It was shown in \cite{Deng_EL2019} that the T-method can be motivated from Bregman's iteration point view. If a blurring filter is the result of solving a variational problem, then the T-method can completely recover the original image.

The R-method, which is originally called the rendition algorithm \cite{milanfar2018rendition}, is developed by minimizing the cost function: $\boldsymbol{x}^T(g(\boldsymbol{x})-\boldsymbol{b})-\frac{1}{2}\boldsymbol{x}^T\boldsymbol{x}$ by using approximations/assumptions and gradient descent. The convergence condition is that $g(.)$ must be Lipschitz continuous \cite{biLipschitz}. The T-method can be considered as a special case of the R-method when $\alpha =1$ and $\beta =1$. This connection provides a gradient descent interpretation of the T-method, which permits applying the accelerated gradient descent algorithms. The R-method can reverse the effect of a wider range of operators than the T-method. However, because of the assumptions and approximations, the R-method only performs well for filters which mildly alter the original image. In a recent paper \cite{delbracio2021polyblur}, similar ideas as that of the R- and T-method are used to develop an algorithm to remove mild defocus and motion blur from natural images. 

The P-method and the S-method \cite{belyaev2021two} are inspired by gradient descent methods studied by Polak \cite{polyak1969minimization} and Steffensen \cite{steffensen1933remarks}. They deliver successful results and converge for a larger number of linear and non-linear filters than other methods. However, they have a high computational cost due to the calculation of the 2-norm of a matrix in each iteration. The 2-norm is the largest singular value of the matrix and has a computational complexity of $O(n^2)$. Because the P-method is more stable than the S-method \cite{belyaev2021two}, we only consider the P-method in this work.

The F-method, which is formulated in the frequency domain \cite{dong2019iterative},  uses the Newton-Raphson technique to invert unknown linear filters. Approximation of the gradient is also used. Although the F-method produces good results for some filters and is reported to have better performance than the T-method and the R-method, it only deals with LSI filters. If the frequency response of the LSI filter has zeros in the frequency range of $[0,\pi]$ then the iteration is unstable. This is a classical problem in inverse filtering, which can be dealt with by using a Wiener filter in the Fourier domain \cite{gonzalez2004digital}.

\begin{table}[h!]

\renewcommand{\arraystretch}{1.25}

    \centering

    \caption{A comparison of the 4 space domain methods, including the proposed TDA-method and a frequency domain method. The complexity is measured in one iteration. The constant $C$ represents the complexity of the filter $g(.)$.\label{tab:compareAlgo} }

   \begin{tabular}{lllll}

\toprule

 \textbf{Method} & \textbf{Update} & \textbf{Para.} & \textbf{Complexity} \tabularnewline

\midrule

T \cite{tao2017zero} & $\boldsymbol{x}_{k+1}=\boldsymbol{x}_{k}+ \boldsymbol{q}_{k}$ & N/A & $\max{(O(n),C)}$ \tabularnewline

R \cite{milanfar2018rendition} & $\boldsymbol{x}_{k+1}=\alpha \boldsymbol{x}_{k}+\lambda \boldsymbol{q}_{k}$ & $\alpha$, $\lambda$ & $\max{(O(n),C)}$\tabularnewline

P  \cite{belyaev2021two} & $\boldsymbol{x}_{k+1}=\boldsymbol{x}_{k}+\frac{||\boldsymbol{q}_{k}||}{2||\boldsymbol{p}_{k}||}\boldsymbol{p}_k$ & N/A & $\max{(O(n^2),C)}$ \tabularnewline

TDA & $\boldsymbol{x}_{k+1}=\boldsymbol{x}_{k}+\lambda \boldsymbol{t}_{k}$ & $\lambda\le1$ & $\max{(O(n),C)}$ \tabularnewline

F \cite{dong2019iterative} & $\mathcal{F}(\boldsymbol{x}_{k+1}) =  \frac{\mathcal{F}(\boldsymbol{b})  }{\mathcal{F}(g(\boldsymbol{x}_k))\mathcal{F}(\boldsymbol{x}_k)} $ & N/A & $\max{(O(n\log n),C)}$ \tabularnewline

\bottomrule 

\end{tabular}

\end{table}

\section{The TDA-method, analysis and accelerations} \label{sec:method}

 \subsection{The proposed TDA-method} \label{sec:TDA}

\begin{figure}[t!]
\centering
\includegraphics[scale=0.4]{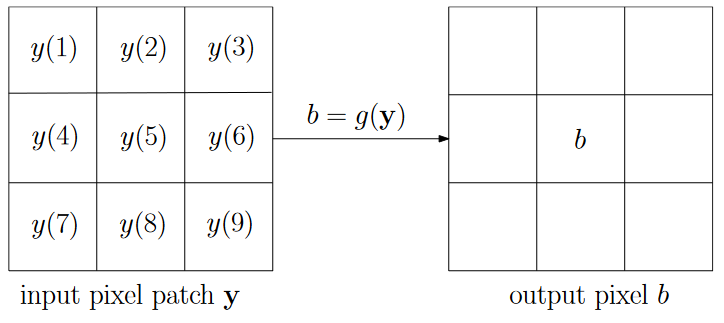}
\caption{\label{fig:A batch processing}Illustration of filtering a $3\times3$ patch resulting a pixel.}
\end{figure}

To develop the proposed algorithm, we assume the filter only acts on a patch of the pixels. For notation simplicity, we represent a patch of pixels as a vector $\boldsymbol{y}$ such that the output pixel is $b=g(\boldsymbol{y})$. We illustrate this in Fig. \ref{fig:A batch processing}. The vector is $\boldsymbol{y}=[y(1),y(2),y(3),y(4),y(5),y(6),y(7),y(8),y(9)]^{T}$ where the pixel to be estimated is $y(5)$ and $b$ is the observed pixel corresponding to $y(5)$. 

We would like to minimize the following cost function to determine $y(5)$:

\begin{equation}
f(\boldsymbol{y})= \frac{1}{2} \left \{ b-g(\boldsymbol{y}) \right \}^{2} \label{eq:j(x)}
\end{equation}
Using gradient descent, we can write

\begin{equation}
\boldsymbol{y}_{k+1}=\boldsymbol{y}_{k}-\lambda\nabla f(\boldsymbol{y}_{k})\label{eq:gd}
\end{equation}
where $\nabla f(\boldsymbol{y})=-u_{k}(5)\boldsymbol{d}$, $u_{k}(5)=b-g(\boldsymbol{y}_{k})$ and $\boldsymbol{d}$ is a vector with its $n$th element defined as $d(n)=\frac{\partial g}{\partial y(n)}$. Because in each patch we are only interested in recovering the center pixel of the patch $y(5)$, from equation (\ref{eq:gd}) we have

\begin{equation}
y_{k+1}(5)=y_{k}(5)+\lambda u_{k}(5)\frac{\partial g}{\partial y(5).}\label{eq:tda1}
\end{equation}

Since the filter function $g$ is unknown, we cannot calculate its derivative and have to resort to approximation. We use the concept of total derivative for the approximation. More specifically, the total derivative of a function $c(\boldsymbol{y}):\mathbb{R}^{N} \rightarrow \mathbb{R}$ at a point $\boldsymbol{a}$ can be approximated as

\begin{equation}
c(\boldsymbol{a}+\boldsymbol{\delta} )-c(\boldsymbol{a})\approx \sum_{n=1}^{N}\delta (n) \frac{\partial c}{\partial y(n)}
\label{eq:td}
\end{equation}
where $y(n)$ and $\delta(n)$ are the $n$th element of the vector $\boldsymbol{y}$  and $\boldsymbol{\delta}$, respectively. With this approximation, we can write the following for the patch: 

\begin{equation}
g(\boldsymbol{y}_{k}+\boldsymbol{u}_{k})-g(\boldsymbol{y}_{k})\approx\sum_{n=1}^{9}u_{k}(n)\frac{\partial g}{\partial y(n)}\label{eq:tda2}
\end{equation} 
where values of elements of $\boldsymbol{u}_k = [u_k(1),u_k(2),...,u_k(9)]$ are assumed to be  small. Except $u_k(5)=b-g(\boldsymbol{y_k})$, other elements are not used in the following approximation. 

A comparison equations (\ref{eq:tda1}) with (\ref{eq:tda2}) suggests a solution of avoiding the calculation of the derivative of an unknown function $\frac{\partial g}{\partial y(5)}$. The solution is through the two-step approximation: 

\begin{equation}
   u_{k}(5)\frac{\partial g}{\partial y(5)} \rightarrow \sum_{n=1}^{9}u_{k}(n)\frac{\partial g}{\partial y(n)} \approx g(\boldsymbol{y}_{k}+\boldsymbol{u}_{k})-g(\boldsymbol{y}_{k}) \label{eq:12}
\end{equation}
where the symbol “$\rightarrow$” represents the operation of “replacing by”. Replacing $u_{k}(5)\frac{\partial g}{\partial y(5)}$ by the local average $\sum_{n=1}^{9}u_{k}(n)\frac{\partial g}{\partial y(n)}$ can be regarded as using a smoothed estimate which is further approximated by the total difference. Using this approximation in equation (\ref{eq:tda1}), we can write 

\begin{equation}
y_{k+1}(5)=y_{k}(5)+\lambda(g(\boldsymbol{y}_{k}+\boldsymbol{u}_{k})-g(\boldsymbol{y}_{k}))\label{eq:tda3}
\end{equation}

We now generalize the above result to the whole image. The key idea is to replace the image patch $\boldsymbol{y}_{k}+\boldsymbol{u}_{k}$ by the image $\boldsymbol{x}_{k}+\boldsymbol{q}_{k}$ where $\boldsymbol{q}_{k}=\boldsymbol{b}-g(\boldsymbol{x}_{k})$. Using this generalization, we can omit the reference to pixel location in (\ref{eq:tda3}) and use our general notation of $\boldsymbol{x}_k$ to represent an image in the $k$th iteration. We can write   
\begin{equation}
\boldsymbol{x}_{k+1}=\boldsymbol{x}_{k}+\lambda(g(\boldsymbol{x}_{k}+\boldsymbol{q}_{k})-g(\boldsymbol{x}_{k}))\label{eq:tda}
\end{equation}
 Equation (\ref{eq:tda}) is the proposed algorithm, which is thus called total derivative approximation (TDA).
\subsection{Convergence condition and image quality} \label{sec:convergence}

It is in general a very difficult task to analyse gradient free algorithms studied in this work, because the large number nonlinear filters make the analysis non-tractable.  In this work, we focus on studying the properties of the T-method and the TDA-method for the mathematically simplest class of filters--LSI low-pass filters. We use frequency domain representation such that $X=\mathcal{F}(\boldsymbol{x})$ and $B=\mathcal{F}(\boldsymbol{b})=\mathcal{F}(g(\boldsymbol{x}))=GX$ where $G$ is the frequency response of the linear filter.

\subsubsection{Frequency domain representations}

We can write the T-method in the frequency domain as follows
\begin{align}
X_{k+1} =X_{k}+B-GX_{k}=B+(1-G)X_{k}\label{eq:FT-method}
\end{align}
After $k$
iterations, we can write the above equation in term of the unknown image $X$ as

\begin{equation}
X_{k}=X-H_{0}^{k}I\label{eq:FTE}
\end{equation}
where $H_{0}=1-G$, $I=(1-G)X$, and  $H_0^k$ represent an element-wise raising to the power of $k$. The result has two terms: the original image $X$ and the result of iteratively filtering the image $(I)$ $k$ times by the same filter $H_{0}$.

For the TDA-method, we perform similar calculations and have the following result:

\begin{equation}
X_{k}=X-J_{0}^{k}I\label{eq:FTDA1}
\end{equation}
where $J_{0}=1-G^{2}$, and $I=(1-G)X$. The result also has two terms: $X$ and the result of iteratively filtering the image $(I)$ $k$ times by using the filter $J_{0}$. In the calculation,  we set $\lambda=1$ such that the TDA-method is consistent with the T-method which is parameter-free.



\subsubsection{Analysis}

From the above results, we can write the two iterative reverse filters in a unified form
\begin{equation}
X_{k}=X-H_{k}I\label{eq:gform}
\end{equation}
where $H_{k}=H_{0}^{k}$ and $H_{0}=1-G$ (T-method), $H_{0}=1-G^{2}$ (TDA-method). In order to recover the original image, a sufficient condition is $|H_{k}|\rightarrow0$ such that $X_{k}\rightarrow X$.

Since $H_{k}$ is a function of the unknown linear filter $G$, to perform a concrete analysis we consider the filter $G$ to be a class of non-causal symmetric linear low pass filters commonly used image processing such as average filters and Gaussian filters whose coefficients sum to 1. The frequency response of such filters, denoted $G(\omega_{1},\omega_{2})$, is real and symmetric and has the property $G(\omega_{1},\omega_{2})\le1$.

\begin{figure}
\begin{centering}
\includegraphics[width=0.5\linewidth]{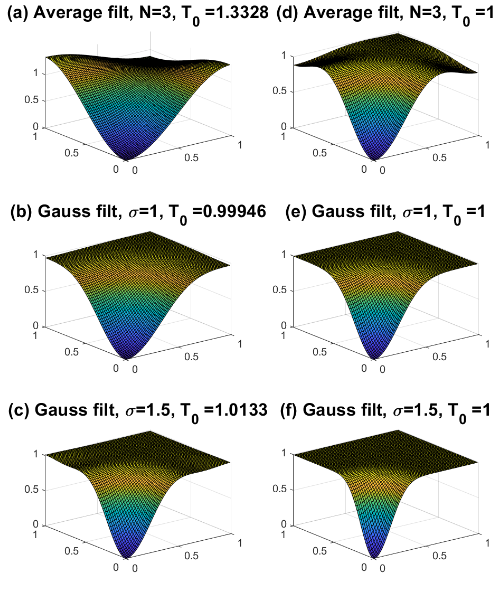}
\par\end{centering}
\caption{\label{fig:The-amplitude-response}The amplitude response $|H_{0}|$
due to 3 linear low pass filters. Left column shows the case for the T-method
where $H_{0}=1-G$. Right column shows the case for the TDA-method
where $H_{0}=1-G^{2}$. The maximum value of $|H_{0}|$ is also shown
as $T_{0}$.}

\end{figure}

Using these notations, the amplitude response of the filter is represented as 
\begin{equation}
|H_{k}(\omega_{1},\omega_{2})|=|H_{0}(\omega_{1},\omega_{2})|^{k}\label{eq:Hk}
\end{equation}
where $H_{0}(\omega_{1},\omega_{2})=1-G(\omega_{1},\omega_{2})$ for the T-method and $H_{0}(\omega_{1},\omega_{2})=1-G^{2}(\omega_{1},\omega_{2})$ for the TDA-method. Let the maximum value of the amplitude response be 
\begin{equation}
T_{0}(\omega_{1}^{*},\omega_{2}^{*})=\max_{(\omega_{1},\omega_{2})}|H_{0}(\omega_{1},\omega_{2})|\label{maxH0}
\end{equation}
where $(\omega_{1}^{*},\omega_{2}^{*})$ represent frequencies where the amplitude response is at maximum. The maximum amplitude response after $k$ iterations, denoted $T_{k}(\omega_{1}^{*},\omega_{2}^{*})$,
is $T_{k}(\omega_{1}^{*},\omega_{2}^{*})=T_{0}^{k}(\omega_{1}^{*},\omega_{2}^{*})$. The super-script $k$ represents raising to the power of $k$. We consider the following three cases.
\begin{itemize}
\item If $T_{0}(\omega_{1}^{*},\omega_{2}^{*})>1$, then $T_{k}(\omega_{1}^{*},\omega_{2}^{*})$ increases exponentially with $k$, which is unbounded. As a result, the frequency component of the image $I(\omega_{1}^{*},\omega_{2}^{*})$
will be amplified by an unbounded factor. The reverse filter is unstable.
\item If $T_{0}(\omega_{1}^{*},\omega_{2}^{*})<1$, then $T_{k}(\omega_{1}^{*},\omega_{2}^{*})$
decreases exponentially with $k$, which approaches the limit of zero. As a result, the amplitude response for all frequencies will approach 0, i.e., $|H_{k}(\omega_{1},\omega_{2})|\rightarrow0$, and the original image is recovered. 
\item If $T_{0}(\omega_{1}^{*},\omega_{2}^{*})=1$, then $|H_{0}(\omega_{1},\omega_{2})|<1$ $(\omega_{1}\ne\omega_{1}^{*},\omega_{2}\ne\omega_{2}^{*})$ and $H_{0}(\omega_{1}^{*},\omega_{2}^{*})=1$.
The amplitude response decreases exponentially $|H_{k}(\omega_{1},\omega_{2})|\rightarrow0$ $(\omega_{1}\ne\omega_{1}^{*},\omega_{2}\ne\omega_{2}^{*})$. As a result, the original image can be approximately recovered except at that particular frequency $(\omega_{1}^{*},\omega_{2}^{*})$.
\end{itemize}
To illustrate the role of $H_{0}(\omega_{1},\omega_{2})$ in determining the behavior of the two reverse filters, we consider three low pass filters: (1) a $(3\times3)$ average filter, (2) a Gaussian low pass filter of standard deviation $\sigma=1$, and (3) a Gaussian low pass filter of standard deviation $\sigma=1.5$.
In Fig. \ref{fig:The-amplitude-response}, we plot $|H_{0}(\omega_{1},\omega_{2})|$ for $0\le\omega_{1},\omega_{2}\le\pi$ (in the figure, the two frequency axes are normalized for easy visualization) for the three filters. The left column shows the results for the T-method, while the right column shows the results of the TDA-method. We have the following observations and explanations.

\begin{itemize}
\item The T-method is unstable for a simple average
filter. This is because the frequency response $G$
can be negative at certain frequency band(s) leading to the amplitude response
$|H_0|=|1-G|>1$ at those frequency band(s). For a
symmetric Gaussian filters, theoretically its frequency response is also a Gaussian, which is always positive under
the assumption that the length of the filter is infinite. However,
in actual implementation the length of the filter is usually set as
$2\times ceil(2\sigma)+1$, where $ceil(x)$ is the ceiling function. The fixed length Gaussian filter can be regarded
as a truncation of the ideal infinite length Gaussian filter. The
truncation can create negative value in $G$ leading to $|H_0|=|1-G|>1$. Experimentally
we found that when $\sigma\le1$, $T_{0}(\omega_{1}^{*},\omega_{2}^{*})\le1$
which results in a stable filter and a prefect recovery of the original
image. When $\sigma>1$, $T_{0}(\omega_{1}^{*},\omega_{2}^{*})>1$
which leads to an unstable filter.
\item The proposed TDA-method is stable for all three filters. This is because
$0\le G^{2}\le1$ and the maximum value of the amplitude response
$|H_0|=|1-G^{2}|\le1$ is attained at frequency where $G^{2}$ is minimum which
is denoted $U_{0}(\omega_{1}^{*},\omega_{2}^{*})=\min_{(\omega_{1},\omega_{2})}G^{2}(\omega_{1},\omega_{2})$.
The maximum value of the amplitude response is thus given by $T_{0}(\omega_{1}^{*},\omega_{2}^{*})=1-U_{0}(\omega_{1}^{*},\omega_{2}^{*})$.
For the three filters, we have $U_{0}(\omega_{1}^{*},\omega_{2}^{*})=0$
leading to $T_{0}(\omega_{1}^{*},\omega_{2}^{*})=1$. 
\item For stable filters, the quality of the recovered image depends on the
range of the frequency $(\omega_{1},\omega_{2})\in\Omega$ such that
$|H_{k}(\omega_{1},\omega_{2})|<1$. More specifically, for the case
$T_{0}(\omega_{1}^{*},\omega_{2}^{*})<1,$ if $\Omega$ covers the whole
frequency plane $[0,\pi]\times[0,\pi]$, then the original image can be
perfectly recovered. For the case $T_{0}(\omega_{1}^{*},\omega_{2}^{*})=1,$  if
$\Omega$ covers more areas of the frequency plane, then the recovered
image is of better quality because more frequency components of the
image $I$ are nulled. In addition, the number of iterations required
to force $|H_{k}(\omega_{1},\omega_{2})|$ to approach zero for frequency $(\omega_{1},\omega_{2})\in\Omega$
 depends on the shape of $|H_{0}(\omega_{1},\omega_{2})|$.
For example, compared to
Fig.\ref{fig:The-amplitude-response} (d) and (e), Fig.\ref{fig:The-amplitude-response}(f) has a larger flat area of the frequency plane satisfying $|H_{0}(\omega_{1},\omega_{2})|\approx1$. As such it will take more iterations to recover an image of better quality for the case of Fig.\ref{fig:The-amplitude-response}(f).
\end{itemize}

\begin{figure}[t]
\begin{centering}
\includegraphics[width=0.9\linewidth]{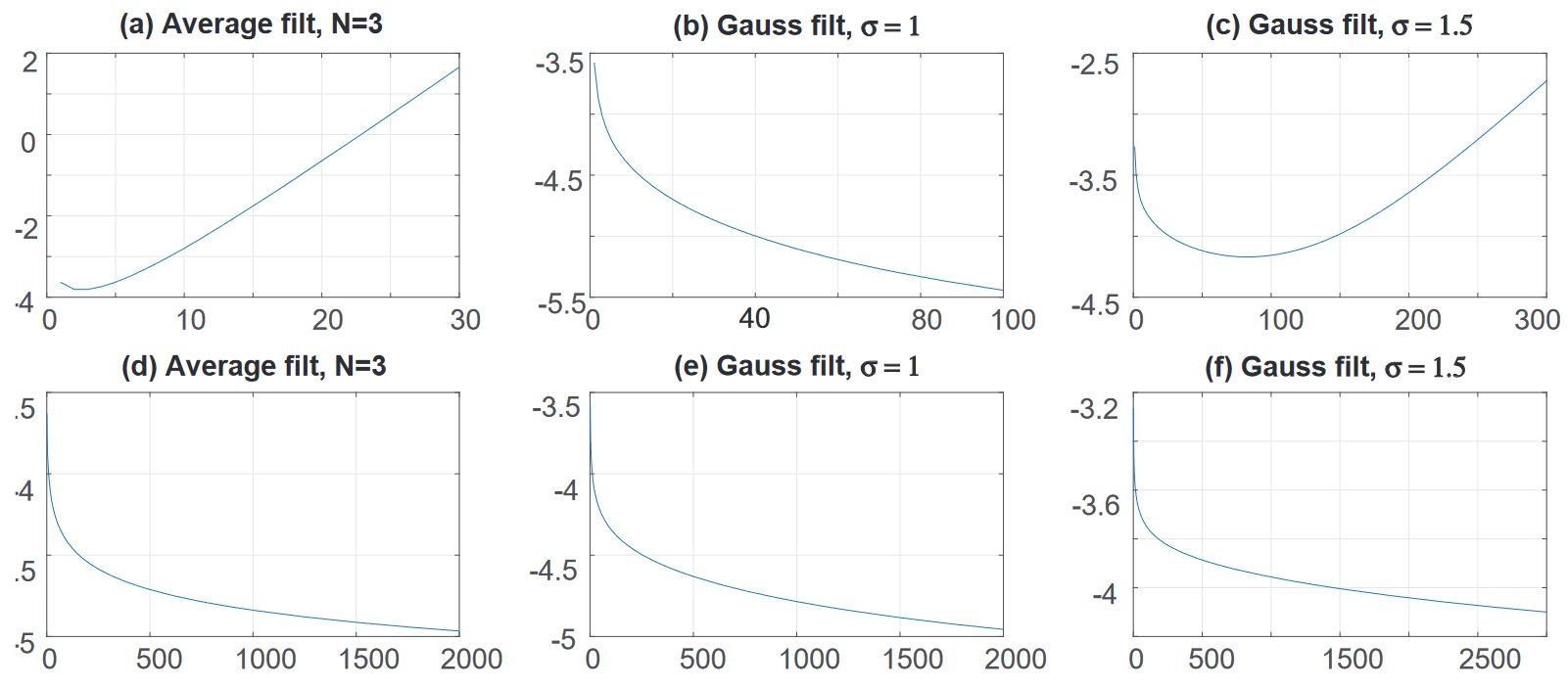}

\par\end{centering}
\caption{\label{fig:The-mean-square}The mean square error (in $\log_{10}$
scale, vertical axis) of the recovered image as a function of number
of iterations (horizontal axis). Top row shows the case for the T-method,
while the bottom row shows the case for the TDA-method.}
\end{figure}

The above analysis is supported by experimental results of an image (the image ``pepper.png’’ in MATLAB) shown in Fig. \ref{fig:The-mean-square}. We used the above 3 linear filters to blur the original image and used the T-method and the proposed TDA-method to recover the original image. We use mean square error (MSE) to measure the quality of the recovered image as a function of the number of iterations. From Fig. \ref{fig:The-mean-square}(a-c), we can see that the T-method is not stable for the average filter and Gaussian filter with $\sigma=1.5$. In fact, this is the case for any average filter and for a Gaussian filter with $\sigma>1$. The T-method can recover the original image perfectly for a Gaussian with $\sigma\le1$, which can be attributed to $T_{0}(\omega_{1}^{*},\omega_{2}^{*})<1$ for this family of filters. 

From Fig. \ref{fig:The-mean-square}(d-f), we can see that the TDA-method is stable for all three filters. This is true for any average filter of any size and a Gaussian filter of any setting of $\sigma$. The required number of iterations to achieve a certain MSE depends on the filter kernel, which determines to what degree the image is smoothed. For example, we can see that for the lightly smoothed image (Fig. \ref{fig:The-mean-square}(d-e) the TDA-method requires about 2000 iterations to achieve a MSE of $10^{-5}$, while for the heavily smoothed image (Fig. \ref{fig:The-mean-square}(f)) the TDA-method only achieves a MSE of $10^{-4}$ at 2000 iterations. Such observation can be further explained in terms of the area of the frequency plane where the amplitude response $|H_{0}(\omega_{1},\omega_{2}|\approx1$. The larger the area such as Fig. \ref{fig:The-amplitude-response}(f), the more iterations are required to recover a good quality approximation of the original image.

\subsection{\label{sec:AGD}Using accelerated gradient descent}

Accelerated gradient descent (AGD) methods \cite{optimizationAlgo},  which are summarised in Table \ref{tab:AGD_table_summary}, were developed to tackle problems related gradient descend. They are widely used training deep neural networks and have the potential to reduce the number of iterations needed to achieve convergence and to increase the robustness against noisy gradients \cite{Goodfellow-et-al-2016}. Since the T-, P- and the proposed TDA-method can be regarded as gradient descent algorithms, we test these AGD methods to see if they can improve the performance of the three algorithms. The tests are conducted by replacing the gradient of each AGD method with the following approximations and results are presented in section \ref{sec:AGD results}.

\begin{itemize}

    \item TDA-method: $\nabla f(\boldsymbol{x_{k}})=-\boldsymbol{t}_{k}$

     \item T-method: $\nabla f(\boldsymbol{x_{k}})=-\boldsymbol{q}_{k}$

    \item P-method: $\nabla f(\boldsymbol{x_{k}}) = -\frac{||\boldsymbol{q}_{k}||}{2||\boldsymbol{p}_{k}||} \boldsymbol{p}_k$

\end{itemize}

\begin{table}
\centering
\caption{AGD methods.}
\label{tab:AGD_table_summary}
\renewcommand{\arraystretch}{1.5}
\begin{tabular}{ll}

\toprule
\textbf{Gradient descent}                        & $\boldsymbol{x}_{k+1}=\boldsymbol{x}_k -\lambda \nabla f(\boldsymbol{x}_k)$                                                                       \vspace{0.25cm}  \\ 
\textbf{MGD}\cite{polyak1964some}                       & $\boldsymbol{x}_{k+1}=\boldsymbol{x}_k +\boldsymbol{v}_k $   \\
 & $ \boldsymbol{v}_k = \beta \boldsymbol{v}_{k-1} -  \lambda \nabla f(\boldsymbol{x}_k)$ \vspace{0.25cm}  \\ 
\textbf{NAG}\cite{nesterov1983method}                     & $\boldsymbol{x}_{k+1}=\boldsymbol{x}_k +\boldsymbol{v}_k $     \\  & $ \boldsymbol{v}_k = \beta \boldsymbol{v}_{k-1} -  \lambda \nabla f(\boldsymbol{x}_k + \beta \boldsymbol{v}_{k-1})$ \vspace{0.25cm}  \\ 

\textbf{RMSProp}\cite{tieleman2012lecture}                     & $\boldsymbol{x}_{k+1}=\boldsymbol{x}_k -\frac{\lambda}{\sqrt{\boldsymbol{v}_k+\epsilon}}\nabla f(\boldsymbol{x}_k)$                                \\ 
                                   & $\boldsymbol{v}_k = \beta \boldsymbol{v}_{k-1} +(1- \beta)\nabla f(\boldsymbol{x}_k)^2$                                                           \vspace{0.25cm}  \\ 
\textbf{Adadelta} \cite{zeiler2012adadelta} & $\boldsymbol{x}_{k+1}=\boldsymbol{x}_k -\Delta \boldsymbol{x}_{k}$ \\
                                    & $\Delta \boldsymbol{x}_{k} =\frac{\sqrt{\boldsymbol{u}_{k-1}+\epsilon}}{\sqrt{\boldsymbol{v}_{k}+\epsilon}}\nabla f(\boldsymbol{x}_k)$                                        \\
                                    & $\boldsymbol{v}_{k} = \beta \boldsymbol{v}_{k-1}  + (1-\beta) \nabla f(\boldsymbol{x}_k)^2$\\
                                   & $\boldsymbol{u}_{k} = \beta \boldsymbol{u}_{k-1}  + (1-\beta) \Delta \boldsymbol{x}_{k}^2$   \vspace{0.25cm}                                                      \\ 
\textbf{ADAM}\cite{kingma2014adam} & $\boldsymbol{x}_{k+1}=\boldsymbol{x}_k -\lambda \frac{\boldsymbol{\hat{m}_k}}{\sqrt{\boldsymbol{\hat{v}_k}}+\epsilon}$                             \\
                                   & $\boldsymbol{m_k} = \beta_1 \boldsymbol{m_{k-1}} + (1-\beta_1) \nabla f(\boldsymbol{x}_k)$  \\
                                   & $\boldsymbol{v_k} = \beta_2 \boldsymbol{v_{k-1}} + (1-\beta_2) \nabla f(\boldsymbol{x}_k)^2$  \\
                                   & $\boldsymbol{\hat{m}_k} = \frac{\boldsymbol{m_k} }{1-\beta_1}$,
                                    $\boldsymbol{\hat{v}_k}= \frac{ \boldsymbol{v_k}}{1-\beta_2}$  \\ \bottomrule                                                                                   
\end{tabular}
\end{table}
\subsection{Summary and comparison}

We can see in Table \ref{tab:compareAlgo} that the T-method is a special case of the R-method and is the simplest of all methods. When the filter $g(.)$ is linear, we have $g(\boldsymbol{x}_{k}+\boldsymbol{q}_{k})-g(\boldsymbol{x}_{k})=g(\boldsymbol{q}_k)$. Thus, compared to the T-method, the TDA-method has a further step of filtering of $\boldsymbol{q}_{k}$ and has a scale factor $\lambda$. The P-method and the TDA-method are related through the different approximations for the gradient.  The computationally expensive ratio of matrix norms in the P-method can be regarded as playing the role of a scale parameter $\lambda$ in the TDA-method. 

\begin{figure}[h!]
    \centering
    \includegraphics[width = 0.5\linewidth]{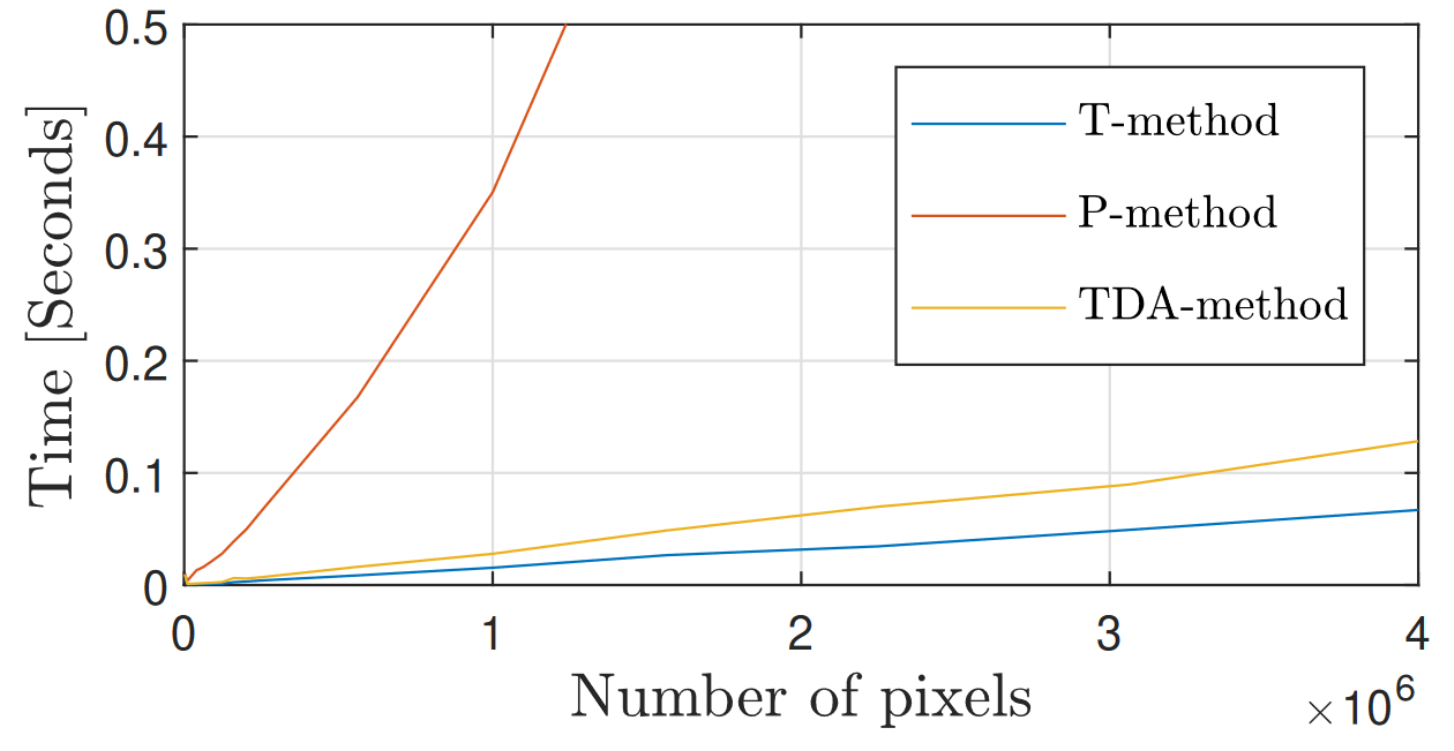}
    \caption{Comparison of running time vs image size when reversing a Gaussian filter with a kernel size of $7\times 7$ pixels and a standard deviation $\sigma=1$ applying one iteration of T-, P- and TDA-method. }
    \label{fig:cc_analysis}
\end{figure}

To illustrate their difference in computational complexity, a MATLAB implementation of the 3 methods is used to reverse a Gaussian filter with a kernel size of $7\times 7$ pixels and $\sigma=1$. Running time for one iteration vs image size is presented in Fig. \ref{fig:cc_analysis}. The P-method is significantly slower than the T- and TDA-method. The running time difference between the T-method and TDA-method is because the latter requires two Gaussian filters per iteration, while the former only requires one.

\begin{figure*}[h!]
\centering
\subfloat[Ground truth]{\includegraphics[width=0.25 \linewidth]{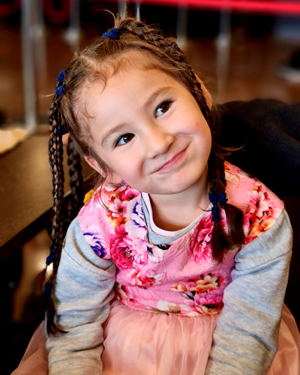}}
\hfil
\subfloat[Smoothed]{\includegraphics[width=0.25 \linewidth]{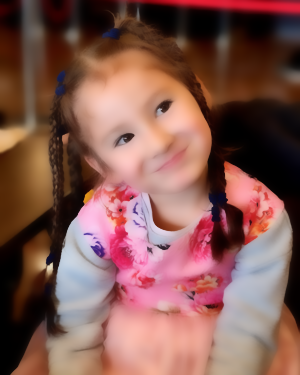}}
\hfil
\subfloat[$N_{iter}=5$]{\includegraphics[width=0.25 \linewidth]{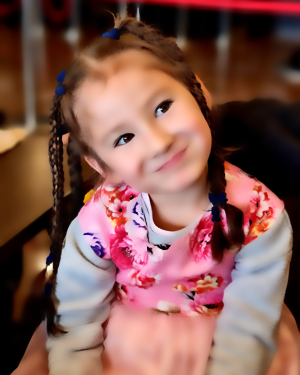}}

\subfloat[$N_{iter}=100$]{\includegraphics[width=0.25 \linewidth]{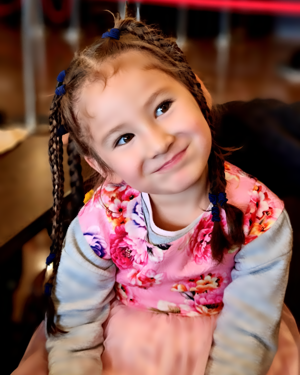}}
\hfil
\subfloat[$N_{iter}=500$]{\includegraphics[width=0.25 \linewidth]{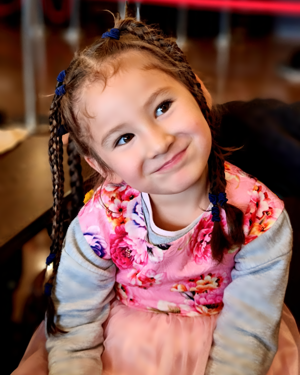}}
\hfil
\subfloat[PSNR vs Iter]{\includegraphics[width=0.25 \linewidth]{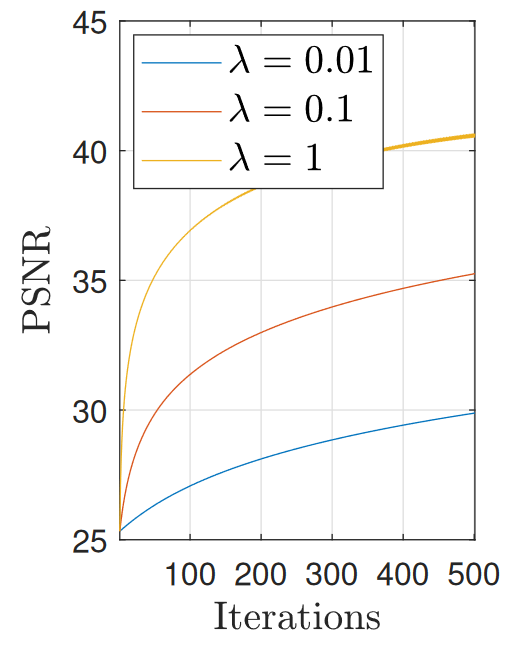}}

\caption{Results of reversing a bilateral filter using the TDA-method with $\lambda = 1$. (a) Original image. (b) Smoothed image with BF (25.3dB). (c) Result after 5 iterations (29.9dB). (d) Result after 100 iterations (37.0dB). (e) Result after 500 iterations (40.9dB). (f) PSNR as a function of $N_{iter}$ for three settings of $\lambda$.}
\label{fig:terations}
\end{figure*}

\section{Experiments and results}

In this section, we present an experimental study of the proposed TDA-method, including effects setting the learning rate $\lambda$ and number of iterations (section \ref{sec:para}), validation of its effectiveness in reversing a wide range of filters (section \ref{sec:applications}), and a comparison (section \ref{sec:comparison}) with of the four existing methods reviewed in section \ref{sec:previous work}. We also present an experimental study (section \ref{sec:AGD results}) of using accelerated gradient descent algorithms. The limitation (section \ref{sec:limitation}) of all current methods is demonstrated through an example in which all of them failed to reverse the effect of a median filter. The running time tests were conducted using MATLAB r2021a running in a PC with Intel i7-3930k CPU and 40GB RAM.

\subsection{Number of iterations, learning rate and image quality}\label{sec:para}
We present two case studies in which the TDA-method is convergent and non-convergent, respectively.
\subsubsection{Convergent case}
We use a successful example of reversing the effect of a bilateral filter (BF) \cite{tomasi1998bilateral} to demonstrate the relationship. Results are shown in Fig. \ref{fig:terations} where the smoothed image is produced by using range and spatial parameters $ \sigma_r=0.3$ and  $\sigma_s=4 $. We can see that proposed TDA-method is convergent and  details from the original image are gradually restored as the number of iterations increases. In this figure, we also show the importance of setting the learning rate $\lambda\le1$. We observed that, for this convergent case, a larger value leads to a faster improvement in image quality.

\subsubsection{Non-convergent case} Although we have shown in last section, the condition for proposed TDA-method to be stable in the linear filter case, the same analysis for a nonlinear filter is in general very difficult. There is no guarantee that the proposed TDA-method, like previous works, will be convergent to a satisfactory result for any filters. Therefore, a practical problem is to determine the criteria for stopping the iteration at a point where the quality of the output image is the highest. 

How do we quantify the output image quality without knowing the original image? Since we have the input image $\boldsymbol{b}=g(\boldsymbol{x})$ and the filter $g(.)$ as a black box, we can calculate the following relative error:

\begin{equation}
    e_k= \frac{||\boldsymbol{b}-g(\boldsymbol{x}_k)||_2^2}{||\boldsymbol{b}||_2^2}
\end{equation}
If $||\boldsymbol{b}-g(\boldsymbol{x}_k)||_2^2$ is small, then under the bi-Lipschitz condition\footnote{We regard the filter $g(.)$ as a function. When it is bi-Lipschitz \cite{biLipschitz},  it satisfies the condition $\frac{1}{K}d_X(x_1,x_2)\le d_Y(y_1,y_2) \le K d_X(x_1,x_2)$ where we define $y=g(x)$, $X$ and $Y$ are set for $x$ and $y$, $K>1$, and $d_X$ and $d_Y$ are metric on sets $X$ and $Y$, respectively. An equivalent definition is that both $g$ and its inverse $g^{-1}$ are Lipschitz.} we can expect that  $||\boldsymbol{x}-\boldsymbol{x}_k||_2^2$  will also be small. Thus, the relative error is a non-referenced metric which can be used to determine when the iteration can be stopped to achieve the best outcome. Indeed, we can use a two-pass process to determine the optimum iterations. In the first pass, we run the algorithm for a fixed number of iterations, record the sequence $\{e_k\}$, and find $n$ such that $n=\min_k \{e_k\}$. In the second pass, we run the algorithm again for $n$ iterations to determine the best outcome. We remark that image quality assessment in general and non-reference quality assessment in particular is an active research area \cite{IQA}. Incorporation results from this research area to deal with the stopping of the iteration is beyond the scope of this work.

In addition, it is a common practice in evaluating image processing algorithms that the original image is assumed to be known so that metrics such as peak-signal-to-noise ratio (PNSR) and structural similarity index (SSIM) can be used. Compared with the relative error, both PSNR and SSIM are referenced metrics which are used to evaluate the performance of the algorithm, especially in comparison of performance of different algorithms. 

\begin{figure*}[h!]
\centering
\subfloat[SSIM]{\includegraphics[width=0.28\linewidth]{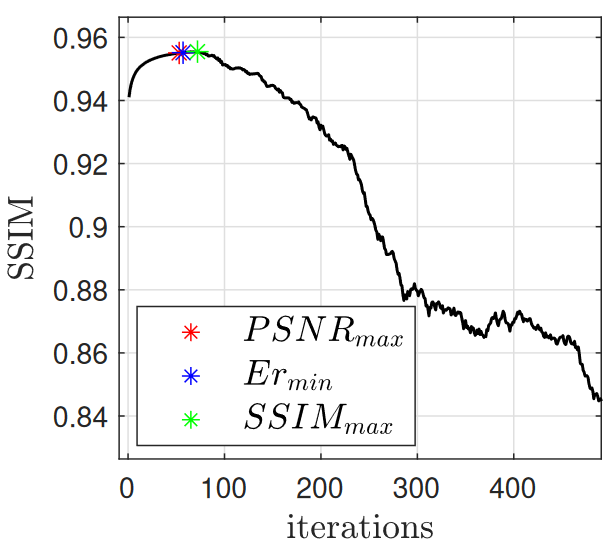}}
\hfil
\subfloat[PSNR]{\includegraphics[width=0.28\linewidth]{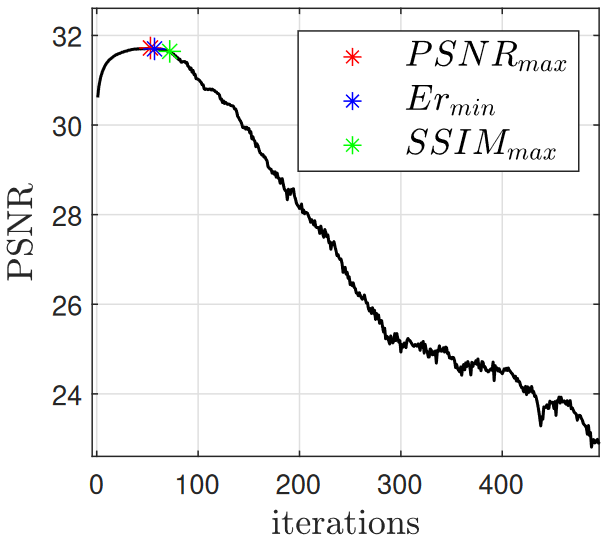}}
\hfil
\subfloat[Relative error]{\includegraphics[width=0.28\linewidth]{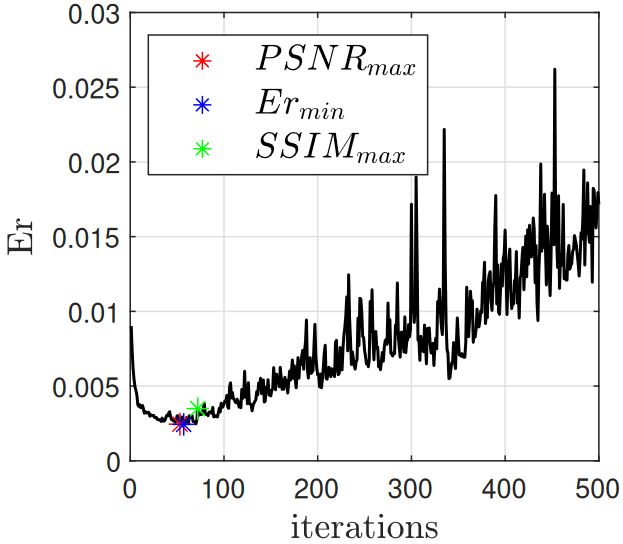}}
\hfil
\subfloat[Original image]{\includegraphics[trim={0 0.5cm 1cm 1.5cm},clip, width=0.22\linewidth]{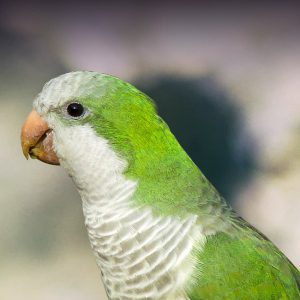}}
\hfil
\subfloat[Input image]{\includegraphics[trim={0 0.5cm 1cm 1.5cm},clip, width=0.22\linewidth]{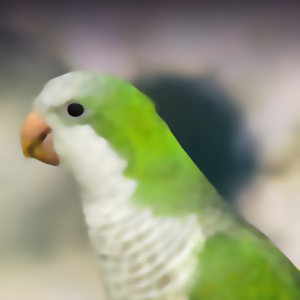}}
\hfil
\subfloat[$N_{iter} = 500$]{\includegraphics[trim={0 0.5cm 1cm 1.5cm},clip, width=0.22\linewidth]{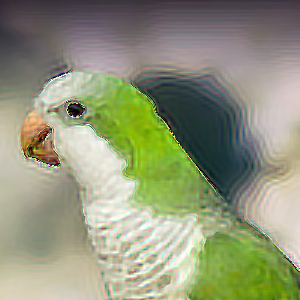}}
\hfil
\subfloat[$N_{iter} = 57$]{\includegraphics[trim={0 0.5cm 1cm 1.5cm},clip, width=0.22\linewidth]{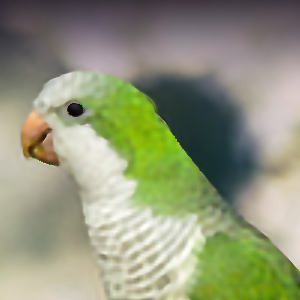}}
\hfil
\caption{Example of finding the optimum number of iterations when reversing a RGF\cite{zhang2014rolling} as non-convergent example with the TDA-method. 
(a)  SSIM per iteration. 
(b) PSNR per iteration. 
(c) Relative error per iteration, (d) original image. 
(e) Smoothed image (30.22dB). (f) Result after 500 iterations (22.93dB). (g) Result after 57 iterations (31.70dB). }
\label{fig:niteroptimum}
\end{figure*}

In Fig.\ref{fig:niteroptimum}, we present a non-convergent example. We use the above 3 metrics in evaluating the performance of reversing a \ac{RGF} \cite{zhang2014rolling} by the TDA-method with $\lambda = 1$. The smoothed image shown in Fig.\ref{fig:niteroptimum}(e) is produced by 4 iterations of a bilateral filter with $\sigma_s=3$ and $\sigma_r=0.05$. We can see that the TDA-method keeps improving output image quality as measured by the three metrics up to a point. After that, the output image quality is getting worse. This is revealed in the top row of the figure, where there is an optimal number of iteration for each metric. We want to point out that the optimal number of iterations determined by the relative error ($k=57$) is fairly close as those determined by using PSNR and SSIM ($k=53$ and $k=72$, respectively). Thus, this experiment support the use of the relative error in determining the number of iterations. In the bottom row of Fig.\ref{fig:niteroptimum}, we can visually compare the results of 500 iterations and 57 iterations. The former (Fig.\ref{fig:niteroptimum}(f)) has strong ringing artifacts, while the latter (Fig.\ref{fig:niteroptimum}(g)) is free of such artifacts and restores an acceptable level of details which can be found in the original image.

\subsubsection{Summary}
When the proposed TDA-method converges, setting a larger value of $\lambda$ helps speed up the convergence rate. When it does not converge to a useful result, we can use the relative error to stop the iteration. Adaptively changing the learning rate will be discussed in section \ref{sec:AGD results}.

\subsection{Evaluation and comparisons}\label{sec:applications}
We first evaluate the effectiveness of the TDA-method by comparing its performance in reversing 12 commonly used filters with that of T- and P-method. We then study three particular cases to highlight and visualize the performance of the TDA-method.

\subsubsection{Evaluation based on 12 filters}\label{sec:Eval}

We performed experiments using 300 natural images from the BSD300 data set \cite{martin2001database}.  We filtered each of them using one of the 12 commonly used image filters listed in Table \ref{tab:filters_parameters}. We then applied the TDA-method of 200 iterations to restore the original image. The average PSNR for each filter in each iteration is calculated over the 300 images. Since different filters produce different levels of degradation, we use the percentage of improvement given by the following equation to compare the performance for different filters 

\begin{equation}
\label{eq:PSNR_rescale}
\bar{p}_k= \frac{p_k - p_0}{p_0}\times 100
\end{equation}
where $p_k$ and $p_0$ represent respectively the average PSNR value for a filter over 300 images at the $k$th iteration and $0$th iteration (the input image). 

\begin{table}[h!]
\centering
\caption{Parameter settings for the 12 filters}
\label{tab:filters_parameters}
\begin{tabular}{l l}
\toprule
\textbf{Filter} & \textbf{Parameters} \\
\midrule
\ac{RGF} \cite{zhang2014rolling} & $\sigma_s=3, \quad \sigma_r=0.05, \quad N_{iter}=4$ \\
Gauss.                                                    & $ \sigma = 5$   \\
\ac{LoG} & $ k = 7 \times 7, \quad \sigma= 0.4 $ \\
\ac{AMF} \cite{gastal2012adaptive}  &   $\sigma_s=7, \quad \sigma_r=0.4$  \\
\ac{RTV} \cite{xu2012structure}     &   $\lambda=0.05, \quad \sigma=3, \quad \epsilon=0.05, \quad N_{iter}=2$     \\
\ac{ILS} \cite{liu2020real} &   $ \lambda= 1,\quad p=0.8,\quad \epsilon=1\times10^{-4},\quad N_{iter}=4 $   \\
\ac{L0} \cite{xu2011image} &   $\lambda = 0.01,\quad \kappa = 2 $    \\
\ac{BF} \cite{tomasi1998bilateral} &    $ \epsilon=0.05,\quad \sigma_s=3 $  \\
Disk averaging  filter             & $ r=3 $       \\
Motion blur filter &     $ l=20,\quad  \theta = 45^{\circ}$   \\
\ac{GF}  \cite{he2012guided}     &  $w_{size}= 5\times5, \quad  \epsilon=0.1$      \\
\ac{GF}+Gauss.       &   $w_{size}= 5\times5, \quad   \epsilon=0.1, \quad  \sigma = 5$     \\

\bottomrule
\end{tabular}
\end{table}

\begin{figure}[h!]
    \centering
    \includegraphics[width = \linewidth]{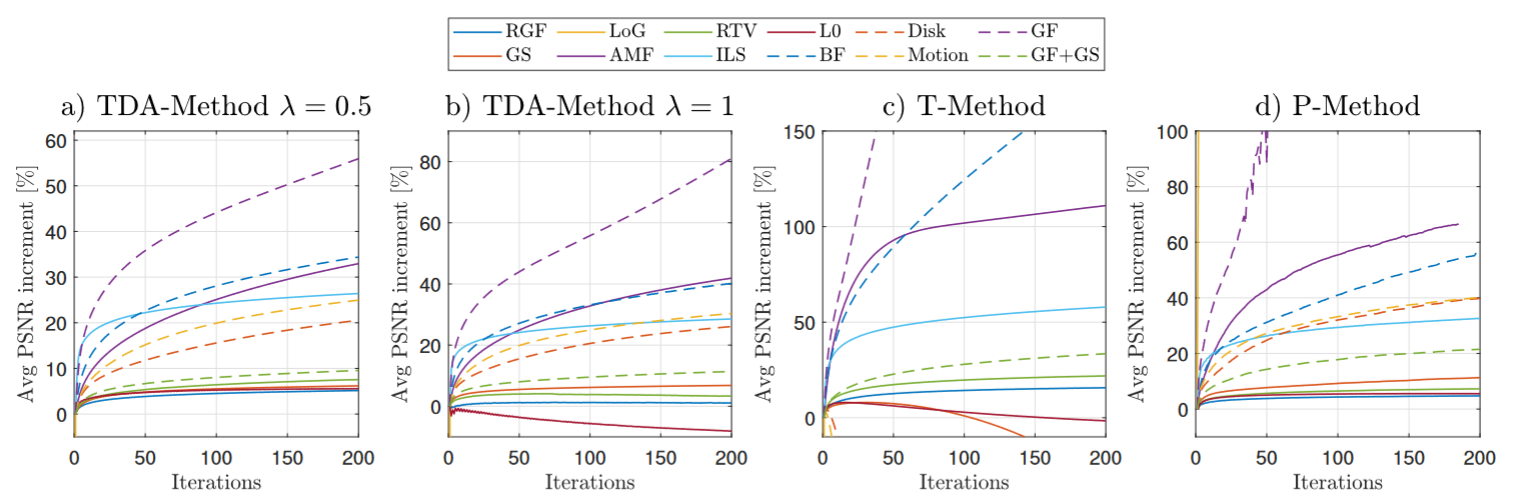}
    \caption{Curves of increment of the Average PSNR per iteration when reversing 12 well known filters using the the BSD300 dataset \cite{martin2001database}. (a) TDA-method setting $\lambda = 0.5$. (b) TDA-method setting $\lambda = 1$. (c) T-method. (d) P-method.}
    \label{fig:BSD300_results}
\end{figure}

The results for the TDA-method are shown in Fig. \ref{fig:BSD300_results}a ($\lambda=0.5$) and Fig. \ref{fig:BSD300_results}b ($\lambda=1$). We can see that the performance of TDA-method depends on the filter being reversed. While it can achieve a substantial improvement for some filters such as GF and BF after a few iterations, it requires a lot of iterations for filters such as Gaussian filter and rolling guided filter. We can also observe that, while setting $\lambda=1$ helps faster improvement in image quality for most filters, such setting does not work in reversing the L0 filter. Overall, after 200 iterations, for 6 out of the 12 filters tested for both settings of $\lambda$, the TDA-method achieved quality improvement greater than 10\%.  For $\lambda=0.5$, improvements have been achieved for all 12 filters after 200 iterations, although some improvements are quite small.

How is the performance of TDA-method when compared to those of the T- and P-method? We conducted the same experiment and presented the results in Fig.\ref{fig:BSD300_results}(c) (T-method)  and (d) (P-method). The T-method produces very good results for 5 filters ($>20\%$ improvement), but it also fails to reverse 5 filters. The P-method, like the TDA-method, is able to reverse all the filters to some extent but at a very high computational cost. Therefore, the proposed TDA method is better than the T-method in terms of its ability to reverse numerous filters, and is also better than the P-method in terms of its much smaller computational cost.

\subsubsection{Subjective and objective comparison examples}\label{sec:comparison}

We present details of three examples for comparing different methods subjectively through visualization and objectively through PSNR. The purpose of the 1st example is to visualize the performance of the TDA-method in reversing two highly nonlinear filters. The 2nd example provides further results of comparing the TDA-method with 4 current methods. The last example demonstrates the effectiveness of the TDA-method in image restoration by comparing it with some classical blind and non-blind algorithms.  

\begin{figure*}[h!]
\centering
\subfloat[Original image]{ \begin{tikzpicture}[
            zoomboxarray,
            zoomboxarray columns=2,
            zoomboxarray rows=1,
            zoombox paths/.append style={line width=0.75pt}
        ]
            \node [image node] { \includegraphics[width=0.15\linewidth]{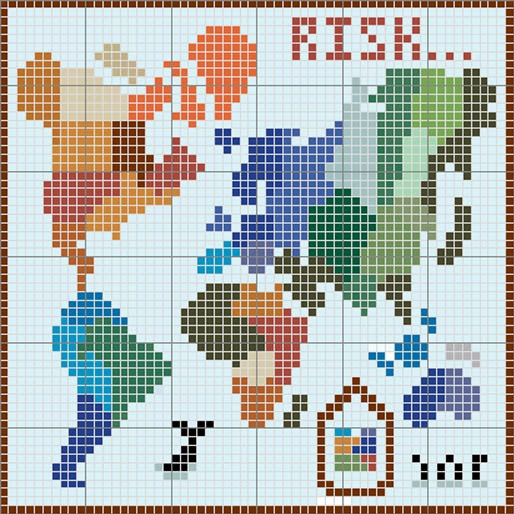} };
            \zoombox[color code = red,magnification=5]{0.3,0.25}  
            \zoombox[color code = blue,magnification=5]{0.45,0.6} 
        \end{tikzpicture}}
        \hfill
\subfloat[RTV (17.87dB)]{\begin{tikzpicture}[
            zoomboxarray,
            zoomboxarray columns=2,
            zoomboxarray rows=1,
            zoombox paths/.append style={line width=0.75pt}
        ]
            \node [image node] { \includegraphics[width=0.15\linewidth]{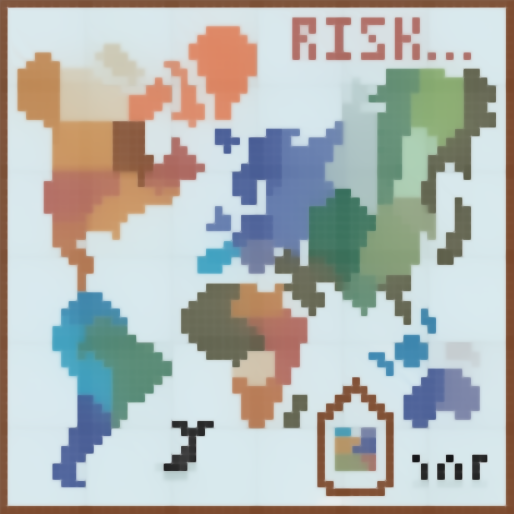} };
            \zoombox[color code = red,magnification=5]{0.3,0.25}  
            \zoombox[color code = blue,magnification=5]{0.45,0.6} 
        \end{tikzpicture}}
        \hfill
\subfloat[Reversed (24.56dB)]{\begin{tikzpicture}[
            zoomboxarray,
            zoomboxarray columns=2,
            zoomboxarray rows=1,
            zoombox paths/.append style={line width=0.75pt}
        ]
            \node [image node] { \includegraphics[width=0.15\linewidth]{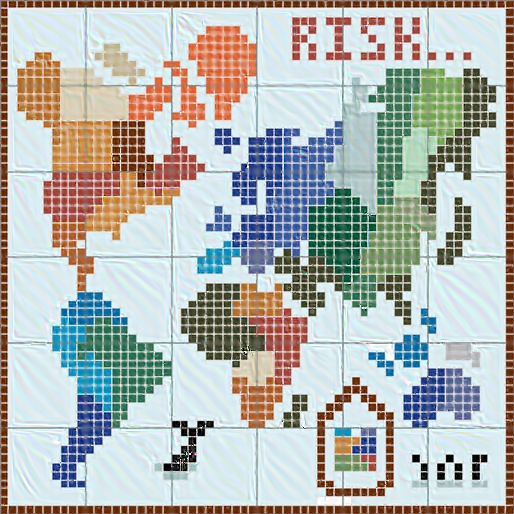} };
            \zoombox[color code = red,magnification=5]{0.3,0.25}  
            \zoombox[color code = blue,magnification=5]{0.45,0.6} 
        \end{tikzpicture}}
        \hfill
        
\subfloat[Original image]{\begin{tikzpicture}[
            zoomboxarray,
            zoomboxarray columns=2,
            zoomboxarray rows=1,
            zoombox paths/.append style={line width=0.75pt}
        ]
            \node [image node] { \includegraphics[width=0.15\linewidth]{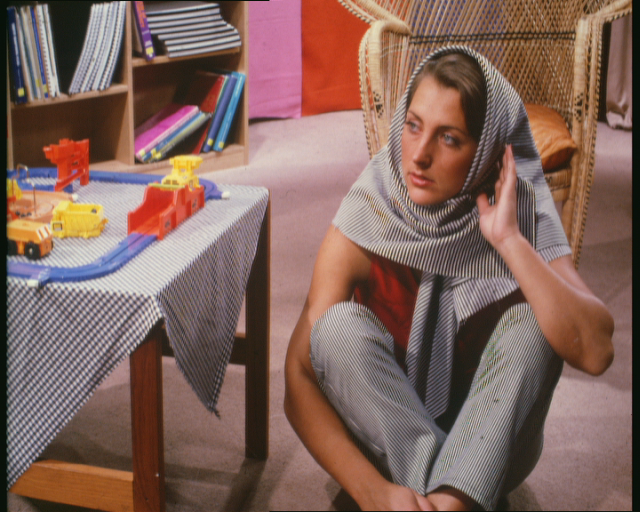} };
            \zoombox[color code = red,magnification=5]{0.3,0.35}  
        \zoombox[color code = blue,magnification=5]{0.7,0.6}  
        \end{tikzpicture}}
        \hfill
\subfloat[SSIF (23.38dB)]{    \begin{tikzpicture}[
            zoomboxarray,
            zoomboxarray columns=2,
            zoomboxarray rows=1,
            zoombox paths/.append style={line width=0.75pt}
        ]
            \node [image node] { \includegraphics[width=0.15\linewidth]{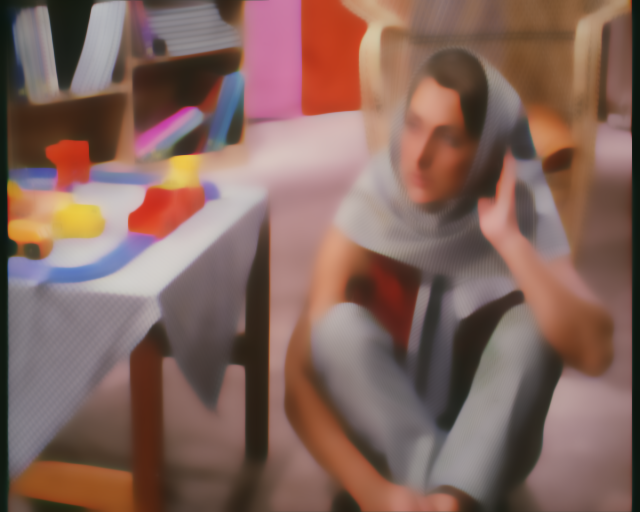} };
            \zoombox[color code = red,magnification=5]{0.3,0.35}  
        \zoombox[color code = blue,magnification=5]{0.7,0.6}  
        \end{tikzpicture}}
        \hfill
\subfloat[Reversed (36.59dB)]{    \begin{tikzpicture}[
            zoomboxarray,
            zoomboxarray columns=2,
            zoomboxarray rows=1,
            zoombox paths/.append style={line width=0.75pt}
        ]
            \node [image node] { \includegraphics[width=0.15\linewidth]{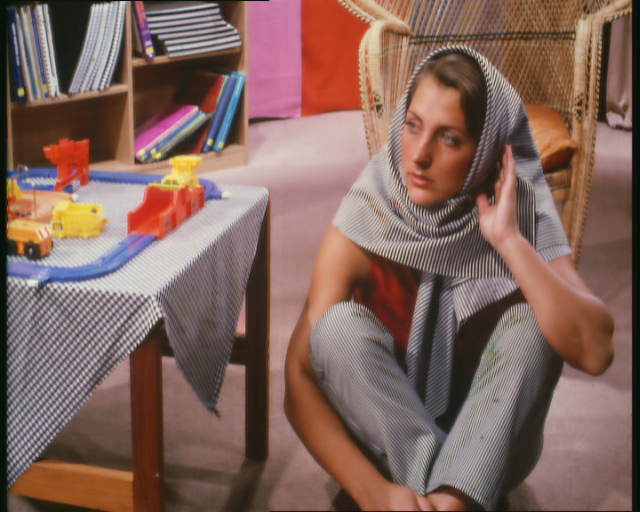} };
            \zoombox[color code = red,magnification=5]{0.3,0.35}  
        \zoombox[color code = blue,magnification=5]{0.7,0.6} 
        \end{tikzpicture}}

\caption{Recovering small-scale image texture by reversing detail removal filters. (a) Original image 1. (b) Filtered image with RTV ($ \lambda = 0.05, \sigma=3, \epsilon = 0.05, N_{iter} = 2 $). (c) Reversed with TDA-method ($\lambda=1, N_{iter} = 3000$). (d) Original image 2. (e) Filtered image with SSIF ($ r = 5, \kappa=0.1, \epsilon = 0.1, N_{iter} = 2 $). (f) Reversed with TDA-method ($\lambda = 1, N_{iter} = 500$). }
\label{fig:smalldetails}
\end{figure*}

\textbf{Example 1}. An image filter can sometimes irreversibly remove details from an image, but in some cases, those details are not completely removed but just diminished at a point of being visually imperceptible. In this example,  we demonstrate the effectiveness of the TDA-method in recovering the texture information which is imperceptible after the image is smoothed by the RTV algorithm \cite{xu2012structure_RTV} or the SSIF algorithm \cite{deng2021guided}. These two algorithms are highly nonlinear and are well known for their ability to smooth out texture. Results are shown in Fig. \ref{fig:smalldetails} which show that the texture information has been recovered to some extent in both images. To produce the results shown in Figures \ref{fig:smalldetails}(c) and (f) we used 3000 and 500 iterations of the TDA-method respectively.

\textbf{Example 2}. We study the performance of the TDA-method in de-convolution applications \cite{gonzalez2004digital}. When the filter kernel is known, the problem is non-blind. Otherwise, the problem is blind. Classical non-blind image restoration algorithms, which require the knowledge of the kernel, include: Lucy Richardson algorithm (LRA) \cite{richardson1972bayesian}, fast image de-convolution using hyper-Laplacian priors (HLP)\cite{krishnan2009fast} and Wiener filter (WF) \cite{wiener1964extrapolation}. A well-known blind restoration algorithm is the maximum likelihood algorithm (MLA) \cite{holmes1995light}, which does not require the knowledge of the kernel. The proposed TDA-method is applicable for a situation where the kernel is unknown but is available as black-box. As such, it can be regarded as semi-blind. We should point out that there is a vast literature on the subject of image restoration, we only pick some classical methods in our experiments.  

\begin{figure*}
\centering
\subfloat[Kernel]{\includegraphics[width=0.24\linewidth,  height = 2.8 cm]{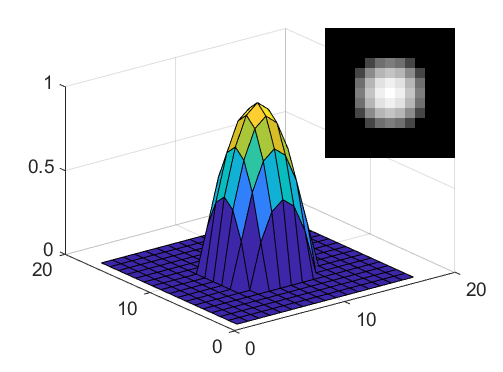}}
\hfil
\subfloat[Ground truth]{\includegraphics[width=0.24\linewidth]{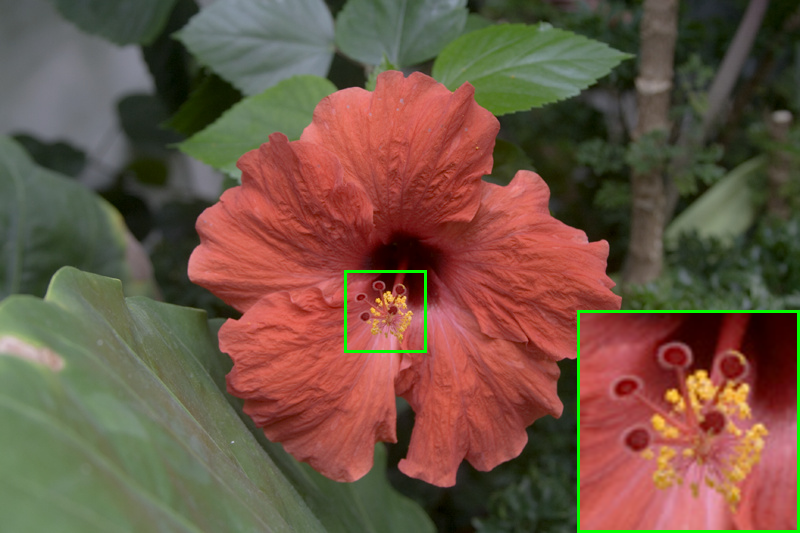}}
\hfil
\subfloat[Filtered (30.02dB)]{\includegraphics[width=0.24\linewidth]{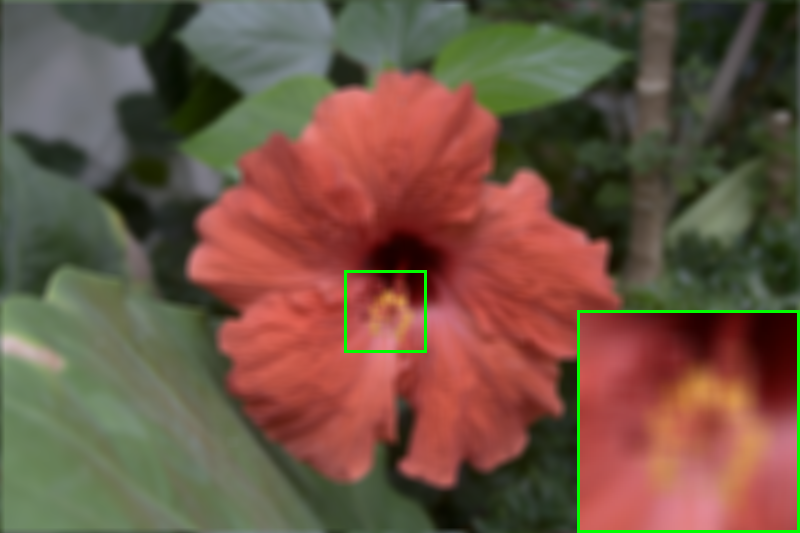}}
\hfil
\subfloat[MLA (33.27dB)\cite{holmes1995light}]{\includegraphics[width=0.24\linewidth]{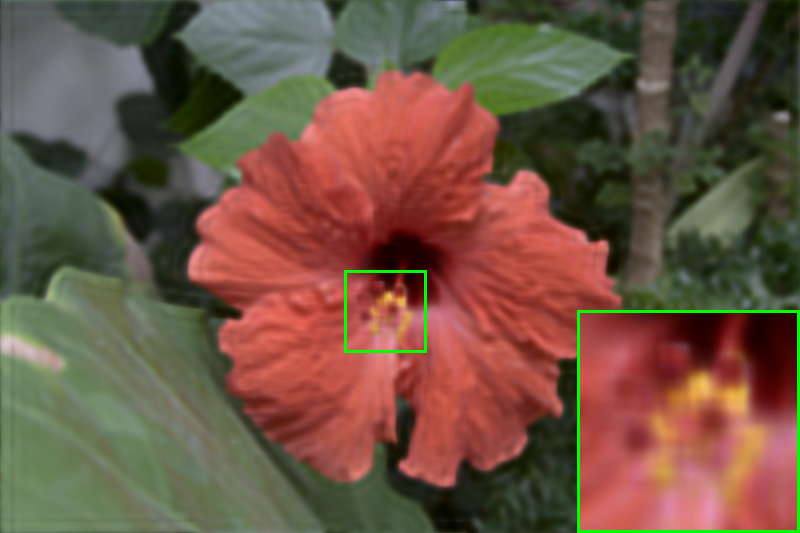}}
\hfil
\subfloat[LRA (34.16dB)\cite{richardson1972bayesian}]{\includegraphics[width=0.24\linewidth]{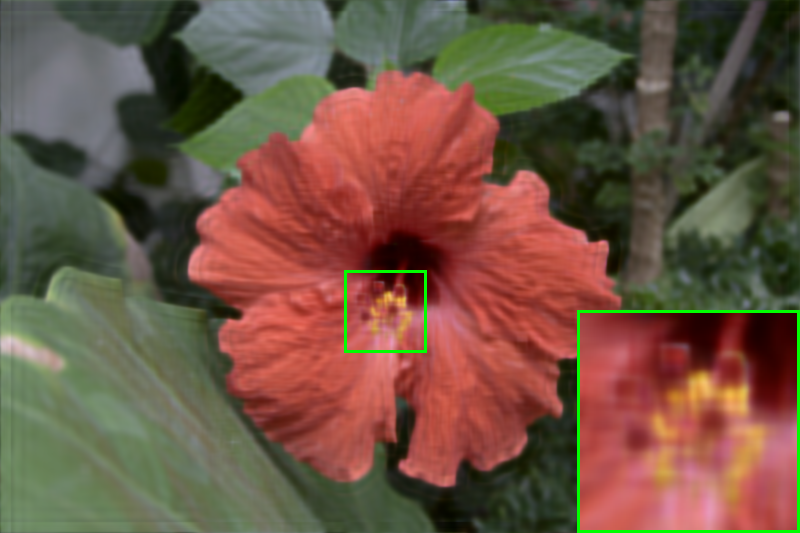}}
\hfil
\subfloat[HLP (33.87dB)\cite{krishnan2009fast}]{\includegraphics[width=0.24\linewidth]{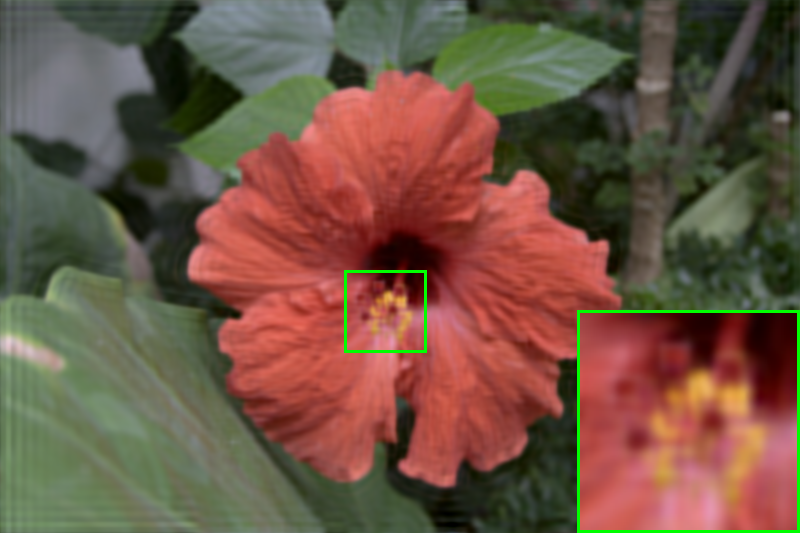}}
\hfil
\subfloat[WF (56.39dB) \cite{wiener1964extrapolation}]{\includegraphics[width=0.24\linewidth]{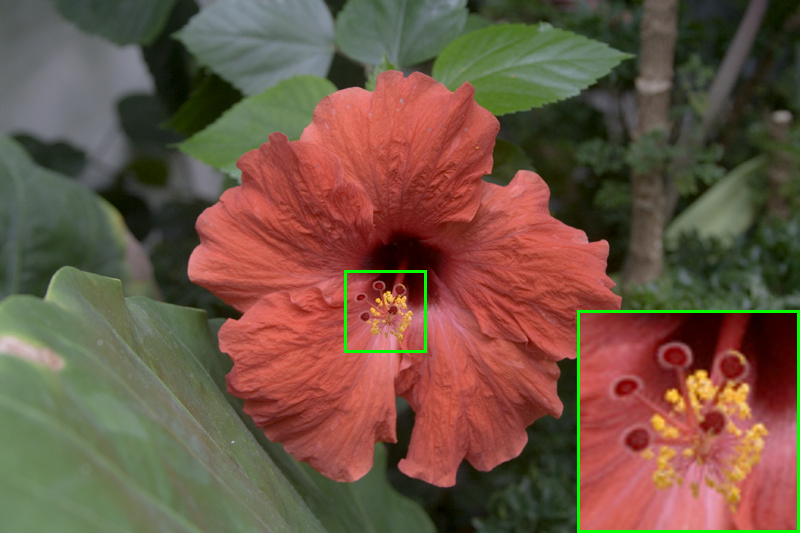}}
\hfil
\subfloat[TDA method (37.67dB)]{\includegraphics[width=0.24\linewidth]{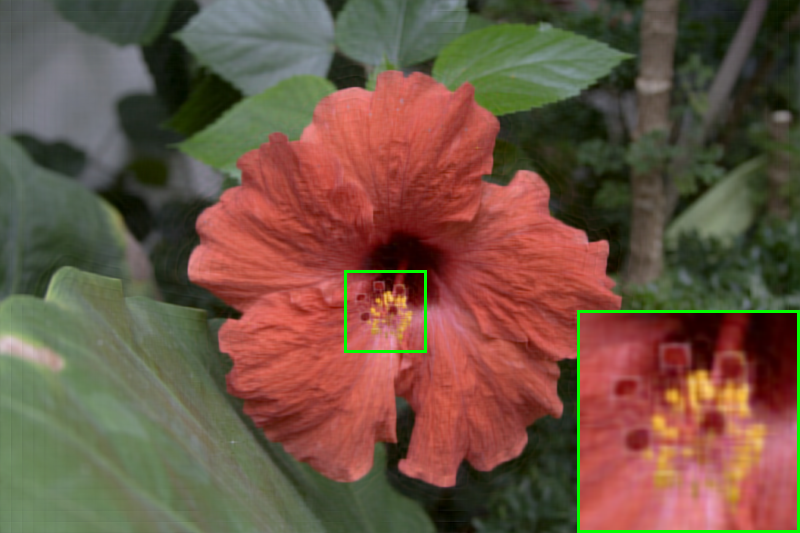}}
\caption{Reversing a Gaussian filter of size $7\times7$ and $\sigma=1$. (a) Kernel. (b) Original image. (c) Filtered image. (d) MLA.  (e) LRA. (f) HLP. (g) Wiener filter.  (h) TDA-method. }
\label{fig:reversingGaussiancomparison}
\end{figure*}

\begin{figure*}
\centering
\subfloat[Kernel]{\includegraphics[width=0.24\linewidth,  height = 2.8 cm]{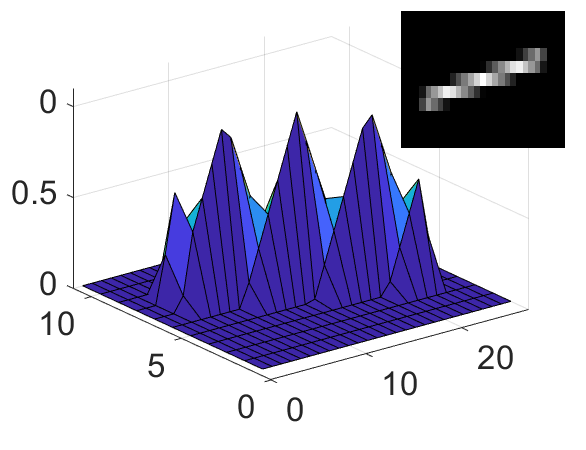}}
\hfil
\subfloat[Ground truth]{\includegraphics[width=0.24\linewidth]{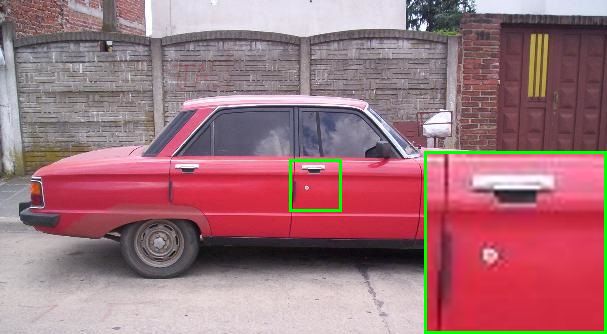}}
\hfil
\subfloat[Filtered (23.45dB)]{\includegraphics[width=0.24\linewidth]{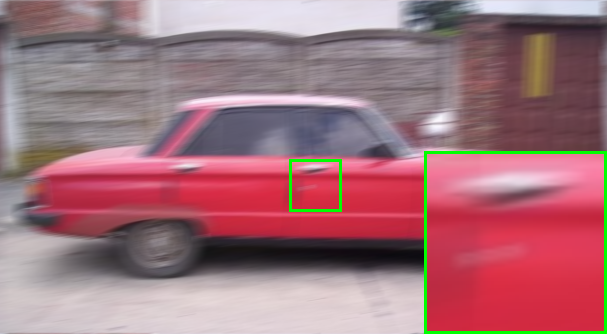}}
\hfil
\subfloat[MLA (26.10dB)\cite{holmes1995light}]{\includegraphics[width=0.24\linewidth]{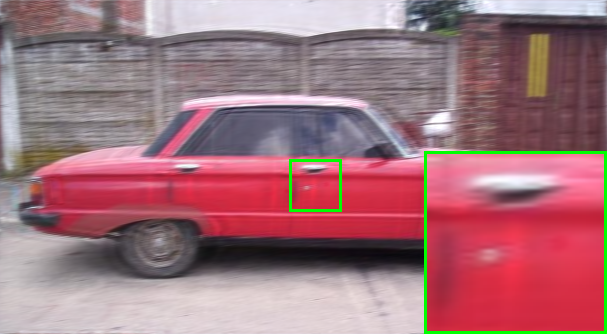}}
\hfil
\subfloat[LRA (27.19dB)\cite{richardson1972bayesian}]{\includegraphics[width=0.24\linewidth]{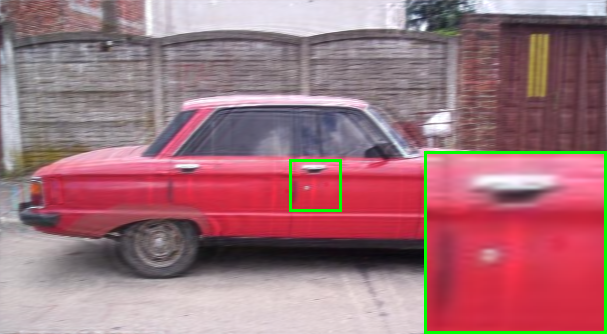}}
\hfil
\subfloat[HLP (27.80dB)\cite{krishnan2009fast}]{\includegraphics[width=0.24\linewidth]{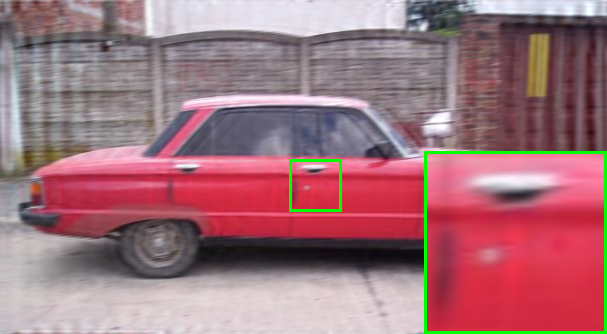}}
\hfil
\subfloat[WF (56.07dB) \cite{wiener1964extrapolation}]{\includegraphics[width=0.24\linewidth]{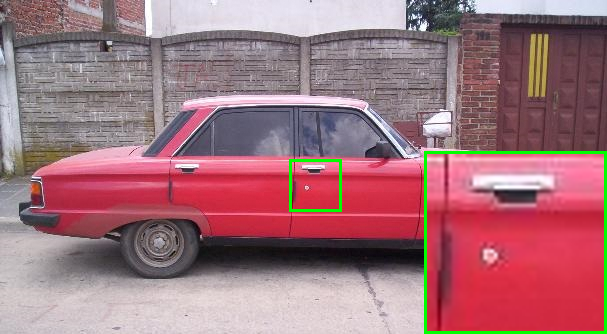}}
\hfil
\subfloat[TDA method (32.17dB)]{\includegraphics[width=0.24\linewidth]{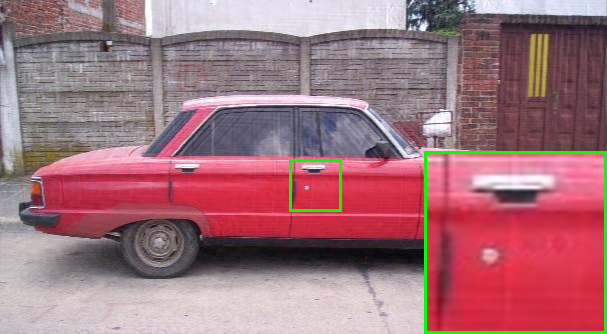}}
\caption{Reversing a motion blur filter. (a) Kernel. (b) Original image. (c) Filtered image. (d) MLA. (e) LRA. (f) HLP. (g) Wiener filter. (h) TDA-method.}
\label{fig:reversingcustomKernelcomparison}
\end{figure*}

In Fig. \ref{fig:reversingGaussiancomparison} we show an example of smoothing using a Gaussian kernel of size $7 \times 7$ pixels and a standard deviation $\sigma = 1$. In Fig. \ref{fig:reversingcustomKernelcomparison} we repeat the same experiment using a kernel which models the blurring due to the linear motion of a camera by 20 pixels with an angle of 10 degrees in a counter-clockwise direction. In both cases, the TDA-method can reverse the effect of the linear filter, small details are recovered and edges are well defined (refer to the green box on Figures  \ref{fig:reversingGaussiancomparison} and \ref{fig:reversingcustomKernelcomparison}). The TDA-method clearly produces a more appealing result than MLA, LRA, and HLP since it does not produce artifacts and the output image is sharper. The Wiener filter is the winner of among all methods. However, it requires the blurring kernel as the input. The impact of not knowing the exact kernel can be seen in the relatively inferior results of the MLA algorithm in both cases. Since the exact knowledge of the kernel may not be available in practice, the proposed method can be useful because it attempts to reverse the effect of any available filter as a black box without the need to know its internal operations.

\begin{figure}[h!]
\centering
\subfloat[Original image]{\includegraphics[width=0.3\linewidth]{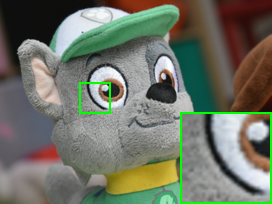}}
\hfil
\subfloat[Filtered (30.27dB)]{\includegraphics[width=0.3\linewidth]{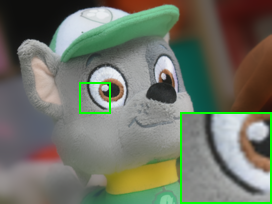}}
\hfil
\subfloat[T-method \cite{tao2017zero} (56.53dB)]{\includegraphics[width=0.3\linewidth]{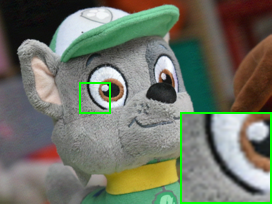}}
\hfil
\subfloat[F-method \cite{dong2019iterative} (43.41dB)]{\includegraphics[width=0.3\linewidth]{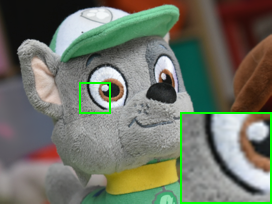}}
\hfil
\subfloat[P-method \cite{belyaev2021two} (56.46dB)]{\includegraphics[width=0.3\linewidth]{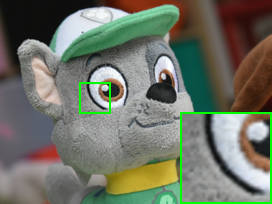}}
\hfil
\subfloat[TDA-method (57.48dB)]{\includegraphics[width=0.3\linewidth]{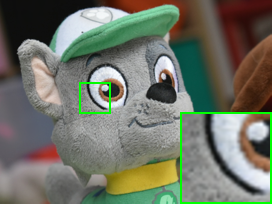}}
\hfil

\caption{Reversing the guided filter \cite{he2012guided}. (a) Original image. (b) Filtered or input image. (c) T-method ($N_{iter} = 10$). (d) F-method ($N_{iter} = 20$). (e) P-method ($N_{iter} = 500$). (f) Proposed TDA-method ($N_{iter} = 500$). }
\label{fig:reversingGFcomparison}
\end{figure}

\textbf{Example 3}. We compare the performance of TDA-method with that of the 4 methods described in section \ref{sec:previous work} in reversing a self-guided filter \cite{he2012guided}. This filter is chosen because it can be reversed by all methods. The filter was configured to produce a texture smoothing effect by setting the window size to $15\times 15$ pixels and $\epsilon = 0.01$. The ground truth and filtered image  are shown in Figures \ref{fig:reversingGFcomparison} (a) and (b), where we can see that textures have been smoothed while the main structure of the image is preserved. Figures \ref{fig:reversingGFcomparison} (c) to (f) show the result for each method. Although the difference between results of all methods is quite small in terms of visual inspection, the TDA-method had achieved the best improvement in PSNR.

\subsection{Results of using accelerated gradient descent methods}\label{sec:AGD results}

In this section, we present a study of applying AGD methods to the three reverse filtering algorithms: the proposed TDA-method, the T-method and the P-method. We summarize the parameter setting of each AGD method in Table \ref{tab:agd_parameters}. We aim to study the following questions.

\begin{table}[h!]
\centering
\caption{AGD methods, parameter settings.}
\label{tab:agd_parameters}
\begin{tabular}{ll}
\toprule
\textbf{Method}   & \textbf{Parameter}                                   \\
\midrule
GD       & $\lambda=1$                                  \\
MGD      & $\lambda=1,\quad  \beta = 0.9,\quad  v_0=0$              \\
NAG      & $\lambda=1,\quad \beta = 0.9,\quad  v_0=0$              \\
RMSprop  & $\lambda=1,\quad \beta = 0.9,\quad  v_0=0$              \\
ADAM     & $\lambda=.1,\quad m_0=0,\quad  v_0=0,\quad  \beta_1=0.9,\quad  \beta_2=0.999$   \\
Adadelta & $\lambda=1,\quad \beta = 0.9,\quad  v_0=0,\quad  u_0=0$        \\
\bottomrule
\end{tabular}
\end{table}

\begin{figure}[h!]
    \centering
        \includegraphics[width =0.65\linewidth]{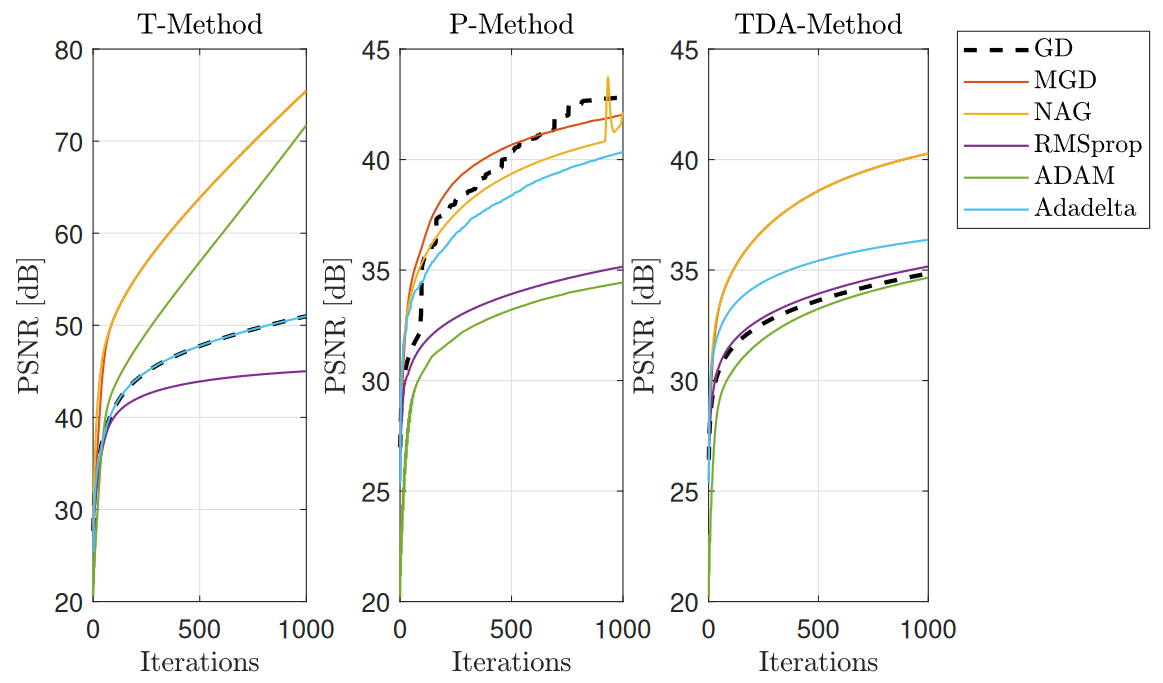}
    \caption{PSNR per iteration when reversing Gaussian filter with T-, P- and TDA-method applying accelerated gradient descent.}
    \label{fig:agd_gaussian_TDA}
\end{figure}

\textbf{Question 1}. When the filter can be reversed, how is the performance of the reverse filter changed by each of AGD method? We conduct an experiment in which all 3 reverse filter methods can successfully recover the original image. We smooth the \textit{"cameraman"} image using a Gaussian filter with a kernel size of $7\times7$ and a standard deviation $\sigma=1$. The three reverse filtering methods and its corresponding AGD variants are then applied to process the image and the PSNR values calculated at each iteration are recorded. Results are presented in Fig. \ref{fig:agd_gaussian_TDA}. Original methods (without AGD and referred to as GD) are represented by thick dashed black lines to simplify the comparison. We can clearly see that for the T-method, the AGD methods of MGD, NAG and ADAM result in significant improvement, while the other two AGD methods, RMSprop and Adadelta do not produce improvement. For the P-method, there is no improvement when using any of the AGD method. Some AGD methods even have a negative impact on its performance. For the TDA-method, both MGD and NAG results in notable improvement, while the other AGD methods lead to roughly the same performance as that of without using AGD.

\begin{figure}[h!]
    \centering
        \includegraphics[width =0.65\linewidth]{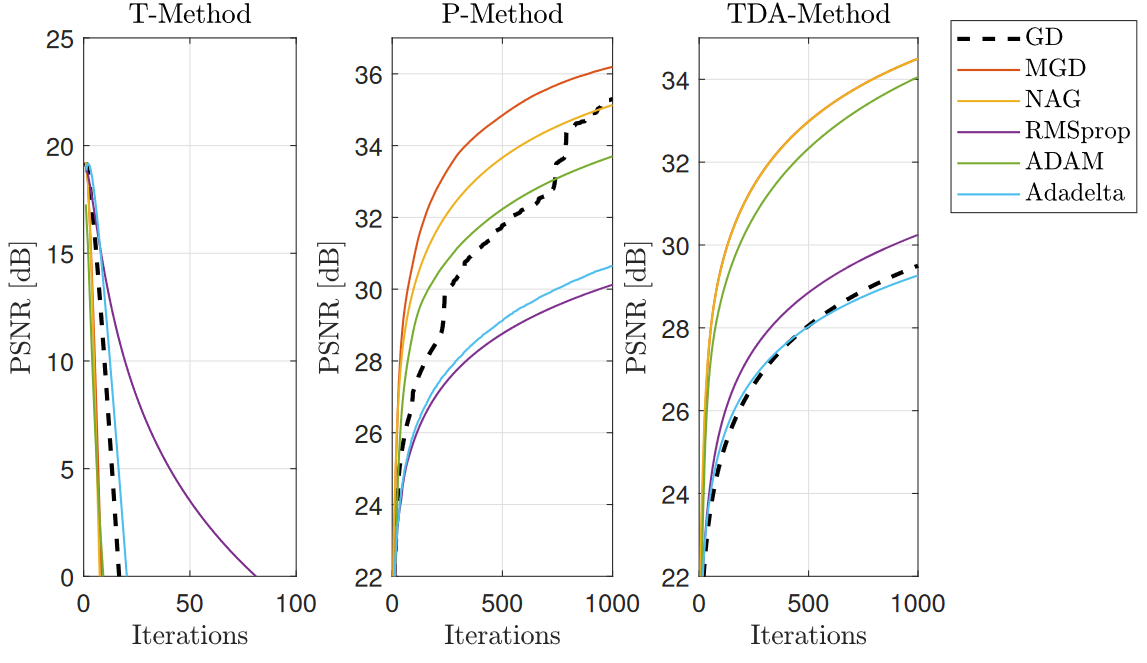}
    \caption{PSNR per iteration when reversing motion blur filter using T-, P- and TDA-method with accelerated gradient descent.}
    \label{fig:agd_MotionBlur}
\end{figure}

\begin{figure}[h!]
\centering
\subfloat[Original]{\includegraphics[trim={0 3cm 0 0},clip, width = 0.32\linewidth]{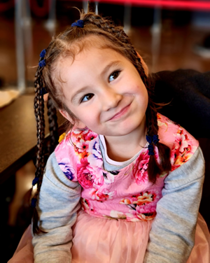}}
\hfil
\subfloat[Blurred (18.84dB)]{\includegraphics[trim={0 3cm 0 0},clip, width =  0.32\linewidth]{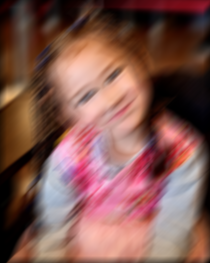}}
\hfil
\subfloat[P without AGD (35.56dB)]{\includegraphics[trim={0 3cm 0 0},clip, width =  0.32\linewidth]{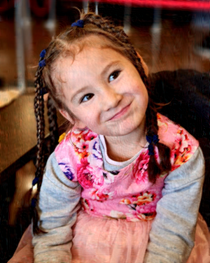}}
\hfil
\subfloat[TDA without AGD (29.71dB)]{\includegraphics[trim={0 3cm 0 0},clip, width =  0.32\linewidth]{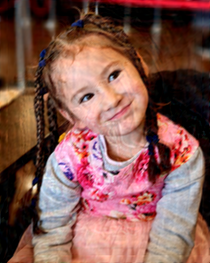}}
\hfil
\subfloat[P with MGD (36.45dB)]{\includegraphics[trim={0 3cm 0 0},clip, width =  0.32\linewidth]{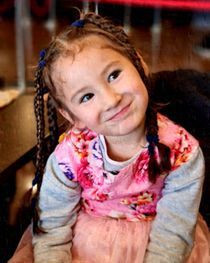}}
\hfil
\subfloat[TDA with NAG (34.76dB)]{\includegraphics[trim={0 3cm 0 0},clip, width =  0.32\linewidth]{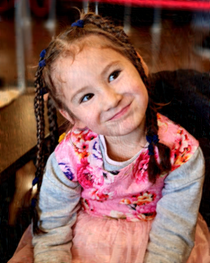}}
\caption{Example of reversing a motion blur filter with P-method and TDA-method by applying accelerated gradient descent. (a) Original image. (b) Blurred image. (c) P-method. (d) TDA-method. (e) P-method with MGD. (f) TDA-method with NAG.}
\label{fig:comparisonmotion}
\end{figure}

\textbf{Question 2}. When the filter cannot be reversed by a particular method because the iteration is an unstable process, does the AGD help to stabilize the iteration?  We conduct an experiment which is aimed at reversing a motion blur filter. An image shown in Fig.  \ref{fig:comparisonmotion}(a) was smoothed by a filter which approximates a linear motion of 20 pixels in an 45 degrees angle. Results are shown in Fig. \ref{fig:agd_MotionBlur}. The T-method fails to recover the image and none of the AGD methods helps to make the T-method produce useful results. 

\textbf{Question 3}. When the T-method failed, how does the AGD help improve the performance of P- and TDA-method? We can see in Fig. \ref{fig:agd_MotionBlur} that performance of the P-method and T-method are improved by using MGD, NAG and ADAM. To provide further evidence of the improvement and to further compare these two methods, we present in Fig. \ref{fig:comparisonmotion} results after 1000 iterations of algorithms with and without AGD. It can be seen that the PSNR improved by about 5dB when the TDA-method is combined with an AGD method, while that of the P-method is improved by about 1dB. However, in 1000 iterations the P-method produces 36.54dB, while the TDA-method produces 34.76dB. We should point out that due to the huge difference in complexity, to complete 1000 iterations, the TDA-method and the P-method take about 3.27 seconds and 43.57 second, respectively. For the TDA-method with NAG to produce the result of 36.5dB, it takes only 14.60 seconds to complete 3048 iterations. This is about 1/3 of the time which the P-method needs to produce the same result. As a further comparison, in 14.60 seconds the P-method could only perform 335 iterations achieving a PSNR of 30.72dB. Therefore, comparing with the P-method, the TDA-method not only benefited more from the AGD, but is also faster due to its much lower complexity.

\textbf{Question 4}. For the three methods, what are the benefits when using AGD over a wide range of filters to be reversed? This is a further study of that presented in \ref{sec:Eval} by comparing the performance of the 3 methods with AGD.  We conducted experiments using the “cameraman” image which is smoothed by 11 filters mentioned in section \ref{sec:Eval}. We have excluded LoG filter because all methods with AGD failed to reverse this filter. We recorded the input PSNR value produced by each filter and the PSNR obtained after applying each reverse filter method after 50 iterations. The results for the TDA-, P- and T-method are shown in Tables \ref{tab:agd_on_filters}, \ref{tab:agd_on_filters_pmethod} and \ref{tab:agd_on_filters_tmethod} respectively, where a negative number indicates a non-convergent iteration. We can see that among the AGD methods  MGD, NAG and Adadelta generally produce the best results. Table \ref{tab:agd_on_filters} shows that, for all smoothing filters, applying AGD to TDA-method results in improved image quality. In addition, Tables \ref{tab:agd_on_filters_pmethod} and \ref{tab:agd_on_filters_tmethod} show that, in some cases, reversing nonlinear filters such as AMF and RTV, the T- and P-method do not benefit from using the AGD. 

\begin{table}
\centering
\caption{PSNR improvement of the TDA-method with AGD techniques. }
\label{tab:agd_on_filters}
\begin{tabular}{lccccccc}
\toprule
\multicolumn{1}{c}{\textbf{Filter}} &
  \multicolumn{1}{c}{\textbf{\rotatebox[origin=c]{60}{\textbf{Input}}}} &
  \multicolumn{1}{c}{\rotatebox[origin=c]{60}{\textbf{GD}}} &
  \multicolumn{1}{c}{\rotatebox[origin=c]{60}{ \textbf{MGD}}} &
  \multicolumn{1}{c}{\rotatebox[origin=c]{60}{\textbf{NAG} }} &
  \multicolumn{1}{c}{\rotatebox[origin=c]{60}{\textbf{RMSprop}}} &
  \multicolumn{1}{c}{\rotatebox[origin=c]{60}{ \textbf{ADAM}}} &
  \multicolumn{1}{c}{\rotatebox[origin=c]{60}{\textbf{Adadelta}}} \\ \midrule
\ac{RGF} \cite{zhang2014rolling}   & \textbf{25} & 23    & -7          & 2           & \textbf{25} & 9  & \textbf{25} \\
Gauss.                            & 25             & 31          & \textbf{33}  & \textbf{33}  & 31               & 29            & 32      \\
\ac{AMF} \cite{gastal2012adaptive} & 20          & 29    & 34          & \textbf{35} & 31          & 33 & 34          \\
\ac{RTV} \cite{xu2012structure}    & 22          & 25    & 24          & 25          & 25          & 23 & \textbf{26} \\
\ac{ILS} \cite{liu2020real}        & 33          & 39    & 38          & 39          & 39          & 33 & \textbf{40} \\
\ac{L0} \cite{xu2011image}         & 27          & 27    & 22          & 23          & \textbf{28} & 24 & 27          \\
\ac{BF} \cite{tomasi1998bilateral} & 28          & 34    & \textbf{39} & 32          & 34          & 35 & 38          \\
Disk       & 22          & 26    & \textbf{30} & 30          & 27          & 28 & 27          \\
Motion                & 19          & 23    & \textbf{27} & \textbf{27} & 24          & 26 & 23          \\
\ac{GF}  \cite{he2012guided}       & 32          & 45    & 47          & \textbf{48}          & 40          & 40 & 47          \\
\ac{GF}+Gauss.                  & 23          & 25    & \textbf{26}          & \textbf{26}          & 25          & 24 & \textbf{26}         \\ \bottomrule
\end{tabular}
\end{table}

\begin{table}[h!]
\centering
\caption{PSNR improvement of the P-method with AGD techniques. }
\label{tab:agd_on_filters_pmethod}
\begin{tabular}{lccccccc}
\toprule
\multicolumn{1}{c}{\textbf{Filter}} &
  \multicolumn{1}{c}{\textbf{\rotatebox[origin=c]{60}{\textbf{Input}}}} &
  \multicolumn{1}{c}{\rotatebox[origin=c]{60}{\textbf{GD}}} &
  \multicolumn{1}{c}{\rotatebox[origin=c]{60}{ \textbf{MGD}}} &
  \multicolumn{1}{c}{\rotatebox[origin=c]{60}{\textbf{NAG} }} &
  \multicolumn{1}{c}{\rotatebox[origin=c]{60}{\textbf{RMSprop}}} &
  \multicolumn{1}{c}{\rotatebox[origin=c]{60}{ \textbf{ADAM}}} &
  \multicolumn{1}{c}{\rotatebox[origin=c]{60}{\textbf{Adadelta}}} \\ \midrule
\ac{RGF} \cite{zhang2014rolling}   & \textbf{25} & \textbf{25} & 0           & 11          & \textbf{25} & 13          & \textbf{25} \\
Gauss.             & 25             & 31          & \textbf{34}  & \textbf{34}  & 31               & 29            & \textbf{34}         \\
\ac{AMF} \cite{gastal2012adaptive} & 20          & \textbf{34} & 33          & 28          & 31          & \textbf{34} & 33          \\
\ac{RTV} \cite{xu2012structure}    & 22          & 25          & 24          & 25          & 25          & 23          & \textbf{26} \\
\ac{ILS} \cite{liu2020real}        & 33          & \textbf{40} & 36          & 39          & 39          & 33          & \textbf{40} \\
\ac{L0} \cite{xu2011image}         & 27          & 27          & 24          & 27          & \textbf{28} & 25          & 27          \\
\ac{BF} \cite{tomasi1998bilateral} & 28          & 34          & 38          & 37          & 34          & 35          & \textbf{40} \\
Disk              & 22          & 27          & \textbf{31} & \textbf{31} & 27          & 28          & 28          \\
Motion                 & 19          & 24          & \textbf{28} & 27          & 25          & 26          & 24          \\
\ac{GF}  \cite{he2012guided}       & 32          & 48          & 47          & \textbf{48}          & 41          & 40          & 47          \\
\ac{GF}+Gauss.                  & 23          & 26          & \textbf{27}          & \textbf{27}          & 25          & 24          & 26         \\ \bottomrule
\end{tabular}
\end{table}

\begin{table}[h!]
\centering
\caption{PSNR improvement of the T-method with AGD techniques. }
\label{tab:agd_on_filters_tmethod}
\begin{tabular}{lccccccc}
\toprule
\multicolumn{1}{c}{\textbf{Filter}} &
  \multicolumn{1}{c}{\textbf{\rotatebox[origin=c]{60}{\textbf{Input}}}} &
  \multicolumn{1}{c}{\rotatebox[origin=c]{60}{\textbf{GD}}} &
  \multicolumn{1}{c}{\rotatebox[origin=c]{60}{ \textbf{MGD}}} &
  \multicolumn{1}{c}{\rotatebox[origin=c]{60}{\textbf{NAG} }} &
  \multicolumn{1}{c}{\rotatebox[origin=c]{60}{\textbf{RMSprop}}} &
  \multicolumn{1}{c}{\rotatebox[origin=c]{60}{ \textbf{ADAM}}} &
  \multicolumn{1}{c}{\rotatebox[origin=c]{60}{\textbf{Adadelta}}} \\ \midrule
\ac{RGF} \cite{zhang2014rolling}   & \textbf{25}          & 14          & -4    & 0           & 14          & 6   & 15          \\
Gauss.                            & 25             & 38          & 46           & \textbf{47}           & 37               & 39   & 38   \\
\ac{AMF} \cite{gastal2012adaptive} & 20          & \textbf{44} & 40    & 42          & 41          & 39  & \textbf{44} \\
\ac{RTV} \cite{xu2012structure}    & 22          & 29          & 29    & \textbf{30} & 29          & 28  & 29          \\
\ac{ILS} \cite{liu2020real}        & 33          & \textbf{43} & 41    & 41          & 41          & 37  & \textbf{43} \\
\ac{L0} \cite{xu2011image}         & 27          & 28          & 25    & 25          & \textbf{29} & 24  & \textbf{29} \\
\ac{BF} \cite{tomasi1998bilateral} & 28          & \textbf{46} & 35    & 35          & 40          & 39  & \textbf{46} \\
Disk             & \textbf{22}          & -14         & -90   & -106        & 5  & -18 & -10         \\
Motion                 & \textbf{19} & -52         & -146  & -176        & 4           & -19 & -40         \\
\ac{GF}  \cite{he2012guided}       & 32          & 53          & 58    & \textbf{73}          & 45          & 43  & 53          \\
\ac{GF}+Gauss.                  & 23          & 29          & \textbf{34}    & \textbf{34}          & 29          & 30  & 29         \\ \bottomrule
\end{tabular}
\end{table}

\subsection{Limitation}\label{sec:limitation}

\begin{figure}[h!]
    \centering
    \includegraphics[width = 0.75 \linewidth]{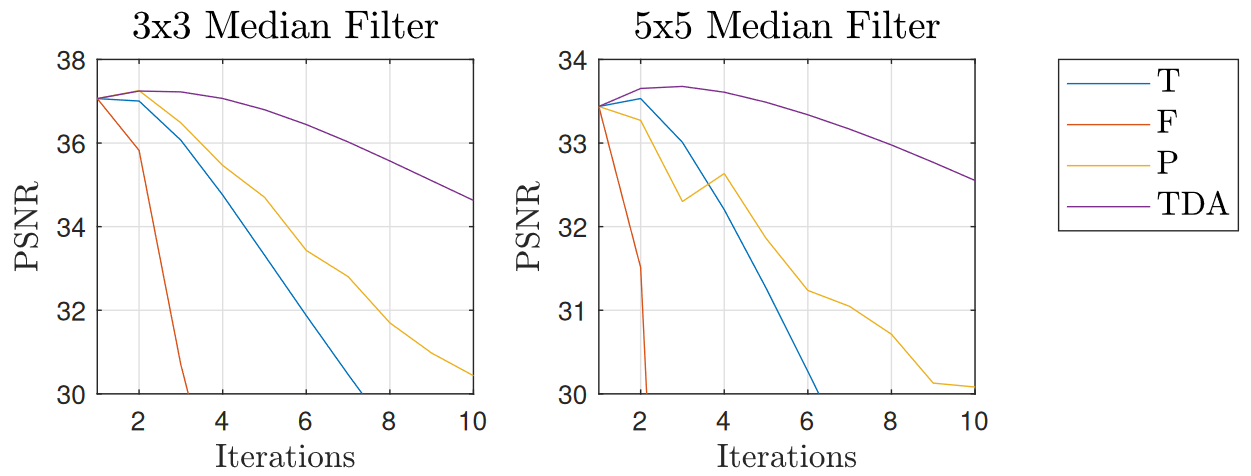}
    \caption{Values of PSNR per iteration obtained when reversing a Median filter using the T-, F-, P- and TDA- methods. (a) 3x3 Median filter. (b) 5x5 Median filter. }
    \label{fig:medianFilter}
\end{figure}

\begin{figure}[h!]
\centering

\subfloat[Original]{\includegraphics[trim={1cm  1.05cmcm 1.4cm 5cm},clip, width = 0.27\linewidth]{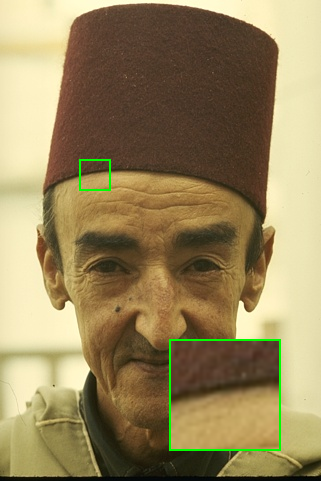}}
\hfil
\subfloat[Filtered (33.43dB)]{\includegraphics[trim={1cm  1.05cm 1.4cm 5cm},clip, width = 0.27\linewidth]{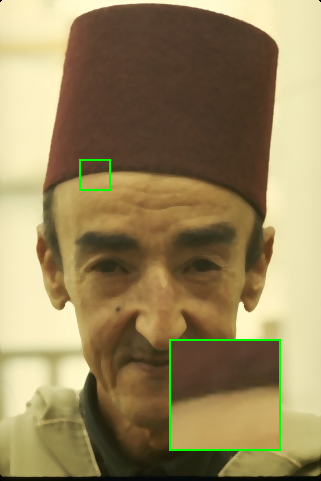}}
\hfil
\subfloat[T-method (25.19dB)]{\includegraphics[trim={1cm  1.05cm 1.4cm 5cm},clip, width = 0.27\linewidth]{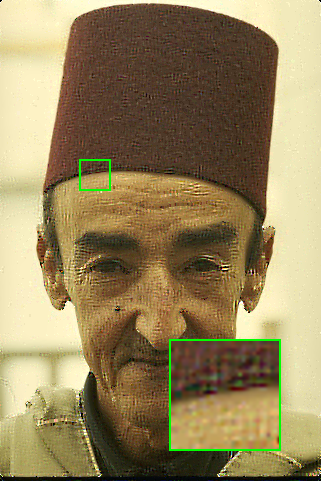}}
\hfil
\subfloat[F-method (6.93dB)]{\includegraphics[trim={1cm  1.05cm 1.4cm 5cm},clip, width = 0.27\linewidth]{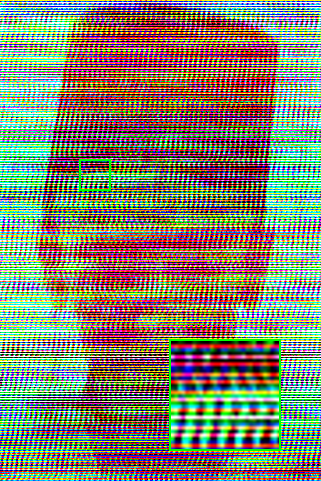}}
\hfil
\subfloat[P-method (29.35dB)]{\includegraphics[trim={1cm  1.05cm 1.4cm 5cm},clip, width = 0.27\linewidth]{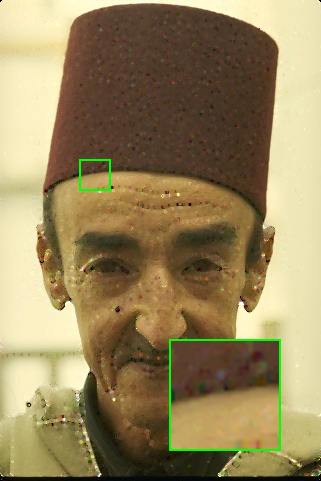}}
\hfil
\subfloat[TDA-method (32.33dB)]{\includegraphics[trim={1cm  1.05cm 1.4cm 5cm},clip, width = 0.27\linewidth]{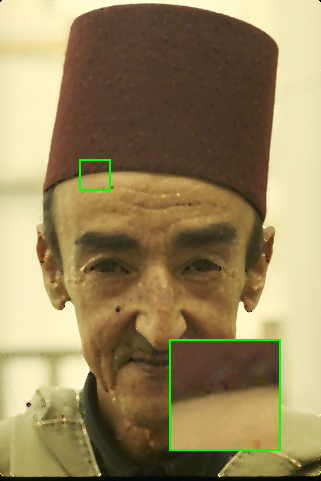}}

\caption{Limitation example. Results of reversing a $5\times5$ median filter using 10 iterations. (a) Original image. (b) Filtered image. (c) T-method. (d) F-method. (e) P-method. (f) TDA-method.}
\label{fig:medianFilter_images}
\end{figure}

We provide a case study in which all algorithms failed to reverse the effect of a simple median filter. We filter an image using a median filter with two patch sizes of $3\times 3$ and $5\times 5$. We then apply 10 iterations of the T-, F-, P- and TDA-method to try to reverse the filter effect.  Results are shown in Figures \ref{fig:medianFilter} and \ref{fig:medianFilter_images}  which show that the PSNR decreases after each iteration and the visual quality of the image is also decreasing.

\newpage
\section{Conclusions}

In this paper, we presented a new solution to a relatively new image inverse problem in which image filter to be reversed is assumed as an available black box. The development of our solution is based on formulating a local patch-based minimization problem and solving it using gradient descent. The problem of unknown gradient due to the unknown filter is solved by using a total derivative based gradient approximation. We study issues related to convergence and output image quality for the case of a linear low pass filter. Results provide new insights into proposed method and the T-method. Because the proposed method and two other existing methods can be regarded as gradient descent algorithms, we study the application of some widely used accelerated gradient descent (AGD) algorithms. We have demonstrated that applying AGD can usually lead to better image quality in smaller number of iterations. However, the success of each AGD method depends on the filter being reversed, so the optimum AGD method needs to be found empirically. 

Through extensive experiments and comparisons with state-of-the-art we have demonstrated that the proposed method has achieved the best trade-off in terms of computational complexity and effectiveness. The proposed method is of the same complexity as the T-method which is the fastest method,  and it is more effective than the T-method in that it can reverse a larger number of filters. The proposed is as effective as that of the P-method in that they can reverse the same set of filters, and it is of much lower complexity than the P-method.

The failure of all gradient-free algorithms to reverse the effect of a median filter calls for further investigation into the assumptions being made in the algorithmic development and possible ways to develop new algorithms which are capable of reversing this family of filters, including order statistics filters. Another possible direction of future work is to combine gradient-free algorithm with  a deep neural network which can reverse the effect of not only one filter, but a wide range of filters. The idea of algorithm unrolling \cite{algorithmUnrolling2021} can be used.   




\begin{thebibliography}{10}

\bibitem{gonzalez2004digital}
R.~Gonzalez, R.~Woods, and S.~Eddins, {\em Digital Image Processing Using
  MATLAB}.
\newblock Pearson Prentice Hall, 2004.

\bibitem{he2010single}
K.~He, J.~Sun, and X.~Tang, ``Single image haze removal using dark channel
  prior,'' {\em IEEE Trans. Pattern anal. Mach. Intell.}, vol.~33, no.~12,
  pp.~2341--2353, 2010.

\bibitem{li2017single}
Z.~Li and J.~Zheng, ``Single image de-hazing using globally guided image
  filtering,'' {\em IEEE Trans. Image Process.}, vol.~27, no.~1, pp.~442--450,
  2017.

\bibitem{Buades2005}
A.~Buades, B.~Coll, and J.~. Morel, ``A non-local algorithm for image
  denoising,'' in {\em Proc. CVPR}, vol.~2, pp.~60--65, IEEE, 2005.

\bibitem{dabov2007image}
K.~Dabov, A.~Foi, V.~Katkovnik, and K.~Egiazarian, ``Image denoising by sparse
  3-d transform-domain collaborative filtering,'' {\em IEEE Trans. Image
  Process.}, vol.~16, no.~8, pp.~2080--2095, 2007.

\bibitem{jain2016survey}
P.~Jain and V.~Tyagi, ``A survey of edge-preserving image denoising methods,''
  {\em Information Systems Frontiers}, vol.~18, no.~1, pp.~159--170, 2016.

\bibitem{tomasi1998bilateral}
C.~{Tomasi} and R.~{Manduchi}, ``Bilateral filtering for gray and color
  images,'' in {\em Proc. ICCV}, pp.~839--846, IEEE, 1998.

\bibitem{he2012guided}
K.~{He}, J.~{Sun}, and X.~{Tang}, ``{Guided image filtering},'' {\em {IEEE
  Trans. Pattern Anal. Mach. Intell.}}, vol.~35, no.~6, pp.~1397--1409, 2013.

\bibitem{wei2018joint}
X.~Wei, Q.~Yang, and Y.~Gong, ``Joint contour filtering,'' {\em Int. J. Comput.
  Vis.}, vol.~126, no.~11, pp.~1245--1265, 2018.

\bibitem{yin2019side}
H.~Yin, Y.~Gong, and G.~Qiu, ``Side window guided filtering,'' {\em Signal
  Process.}, vol.~165, pp.~315--330, 2019.

\bibitem{Sun2020}
Z.~Sun, B.~Han, J.~Li, J.~Zhang, and X.~Gao, ``Weighted guided image filtering
  with steering kernel,'' {\em IEEE Trans. Image Process.}, vol.~29,
  pp.~500--508, 2020.

\bibitem{li2015weighted}
Z.~Li, J.~Zheng, Z.~Zhu, W.~Yao, and S.~Wu, ``Weighted guided image
  filtering,'' {\em IEEE Trans. Image Process.}, vol.~24, no.~1, pp.~120--129,
  2014.

\bibitem{endoSA2017}
Y.~Endo, Y.~Kanamori, and J.~Mitani, ``Deep reverse tone mapping,'' {\em ACM
  Trans. Graph. (Proc. of SIGGRAPH ASIA 2017)}, vol.~36, Nov. 2017.

\bibitem{CNN_deblur2021}
J.~Koh, J.~Lee, and S.~Yoon, ``Single-image deblurring with neural networks: A
  comparative survey,'' {\em Comput. Vis. Image Underst.}, vol.~203, p.~103134,
  2021.

\bibitem{CT1988}
A.~C. Kak and M.~Slaney, {\em Principles of Computerized Tomographic Imaging}.
\newblock IEEE Press, 1988.

\bibitem{CompressedSensing}
J.~Romberg, ``Imaging via compressive sampling,'' {\em IEEE Signal Process.
  Mag.}, vol.~25, no.~2, pp.~14--20, 2008.

\bibitem{tao2017zero}
X.~Tao, C.~Zhou, X.~Shen, J.~Wang, and J.~Jia, ``Zero-order reverse
  filtering,'' in {\em Proc. ICCV}, pp.~222--230, IEEE, 2017.

\bibitem{milanfar2018rendition}
P.~Milanfar, ``Rendition: Reclaiming what a black box takes away,'' {\em arXiv
  preprint arXiv:1804.08651}, 2018.

\bibitem{belyaev2021two}
A.~G. Belyaev and P.-A. Fayolle, ``Two iterative methods for reverse image
  filtering,'' {\em Signal, Image and Video Process.}, pp.~1--9, 2021.

\bibitem{dong2019iterative}
L.~Dong, J.~Zhou, C.~Zou, and Y.~Wang, ``Iterative first-order reverse image
  filtering,'' in {\em Proceedings of the ACM Turing Celebration
  Conference-China}, pp.~1--5, 2019.

\bibitem{zhang2014rolling}
Q.~Zhang, X.~Shen, L.~Xu, and J.~Jia, ``Rolling guidance filter,'' in {\em
  Proc. ECCV}, pp.~815--830, Springer, 2014.

\bibitem{gastal2012adaptive}
E.~S. Gastal and M.~M. Oliveira, ``Adaptive manifolds for real-time
  high-dimensional filtering,'' {\em ACM Trans. Graph.}, vol.~31, no.~4,
  pp.~1--13, 2012.

\bibitem{xu2012structure}
L.~Xu, Q.~Yan, Y.~Xia, and J.~Jia, ``Structure extraction from texture via
  relative total variation,'' {\em {ACM Trans. Graph.}}, vol.~31, no.~6,
  pp.~1--10, 2012.

\bibitem{liu2020real}
W.~Liu, P.~Zhang, X.~Huang, J.~Yang, C.~Shen, and I.~Reid, ``Real-time image
  smoothing via iterative least squares,'' {\em ACM Trans. Graph.}, vol.~39,
  no.~3, pp.~1--24, 2020.

\bibitem{xu2011image}
L.~Xu, C.~Lu, Y.~Xu, and J.~Jia, ``{Image smoothing via $L_0$ gradient
  minimization},'' {\em ACM Trans. Graph.}, vol.~30, no.~6, pp.~1--12, 2011.

\bibitem{optimizationAlgo}
M.~J. Kochenderfer and T.~A. Wheeler, {\em Algorithms for Optimization}.
\newblock The {MIT} Press, 2019.

\bibitem{Goodfellow-et-al-2016}
I.~Goodfellow, Y.~Bengio, and A.~Courville, {\em Deep Learning}.
\newblock The MIT Press, 2016.

\bibitem{ortega2000iterative}
J.~M. Ortega and W.~C. Rheinboldt, {\em Iterative Solution of Nonlinear
  Equations in Several Variables}.
\newblock SIAM, 2000.

\bibitem{Deng_EL2019}
G.~Deng and P.~Broadbridge, ``Bregman inverse filter,'' {\em Electron. Lett.},
  vol.~55, no.~4, pp.~192--194, 2019.

\bibitem{biLipschitz}
N.~Weaver, {\em Lipschitz Algebras}.
\newblock World Scientific Publishing {C}o. {P}te. {L}td., 2018.

\bibitem{delbracio2021polyblur}
M.~Delbracio, I.~Garcia-Dorado, S.~Choi, D.~Kelly, and P.~Milanfar, ``Polyblur:
  Removing mild blur by polynomial reblurring,'' {\em IEEE Trans. Comput.
  Imaging}, vol.~7, pp.~837--848, 2021.

\bibitem{polyak1969minimization}
B.~T. Polyak, ``Minimization of unsmooth functionals,'' {\em USSR Computational
  Mathematics and Mathematical Physics}, vol.~9, no.~3, pp.~14--29, 1969.

\bibitem{steffensen1933remarks}
J.~Steffensen, ``Remarks on iteration,'' {\em Scandinavian Actuarial Journal},
  vol.~1933, no.~1, pp.~64--72, 1933.

\bibitem{polyak1964some}
B.~T. Polyak, ``Some methods of speeding up the convergence of iteration
  methods,'' {\em Ussr Computational Mathematics and Mathematical Physics},
  vol.~4, no.~5, pp.~1--17, 1964.

\bibitem{nesterov1983method}
Y.~E. Nesterov, ``A method for solving the convex programming problem with
  convergence rate o (1/k\^{} 2),'' in {\em Dokl. Akad. Nauk Sssr}, vol.~269,
  pp.~543--547, 1983.

\bibitem{tieleman2012lecture}
T.~Tieleman and G.~Hinton, ``Lecture 6.5-rmsprop, coursera: Neural networks for
  machine learning,'' {\em University of Toronto, Technical Report}, 2012.

\bibitem{zeiler2012adadelta}
M.~D. Zeiler, ``Adadelta: an adaptive learning rate method,'' {\em arXiv
  preprint arXiv:1212.5701}, 2012.

\bibitem{kingma2014adam}
D.~P. Kingma and J.~Ba, ``Adam: A method for stochastic optimization,'' {\em
  arXiv preprint arXiv:1412.6980}, 2014.

\bibitem{IQA}
G.~Zhai and X.~Min, ``Perceptual image quality assessment: a survey,'' {\em
  Sci. China Inf. Sci.}, vol.~63, pp.~211301:1--211301:52, 2020.

\bibitem{martin2001database}
D.~Martin, C.~Fowlkes, D.~Tal, and J.~Malik, ``A database of human segmented
  natural images and its application to evaluating segmentation algorithms and
  measuring ecological statistics,'' in {\em Proc. ICCV}, vol.~2, pp.~416--423,
  IEEE, 2001.

\bibitem{xu2012structure_RTV}
L.~Xu, Q.~Yan, Y.~Xia, and J.~Jia, ``Structure extraction from texture via
  relative total variation,'' {\em ACM Trans. Graph. (TOG)}, vol.~31, no.~6,
  pp.~1--10, 2012.

\bibitem{deng2021guided}
G.~Deng, F.~Galetto, M.~Al-nasrawi, and W.~Waheed, ``A guided edge-aware
  smoothing-sharpening filter based on patch interpolation model and
  generalized gamma distribution,'' {\em IEEE Open Journal of Signal Process.},
  vol.~2, pp.~119--135, 2021.

\bibitem{richardson1972bayesian}
W.~H. Richardson, ``Bayesian-based iterative method of image restoration,''
  {\em J. Acoust. Soc. Am.}, vol.~62, no.~1, pp.~55--59, 1972.

\bibitem{krishnan2009fast}
D.~Krishnan and R.~Fergus, ``Fast image deconvolution using hyper-laplacian
  priors,'' {\em Advances in Neural Information Processing Systems}, vol.~22,
  pp.~1033--1041, 2009.

\bibitem{wiener1964extrapolation}
N.~Wiener {\em et~al.}, {\em Extrapolation, interpolation, and smoothing of
  stationary time series: with engineering applications}, vol.~8.
\newblock The MIT Press, 1964.

\bibitem{holmes1995light}
T.~J. Holmes, S.~Bhattacharyya, J.~A. Cooper, D.~Hanzel, V.~Krishnamurthi,
  W.-C. Lin, B.~Roysam, D.~H. Szarowski, and J.~N. Turner, ``Light microscopic
  images reconstructed by maximum likelihood deconvolution,'' in {\em Handbook
  of Biological Confocal Microscopy}, pp.~389--402, Springer, 1995.

\bibitem{algorithmUnrolling2021}
V.~Monga, Y.~Li, and Y.~C. Eldar, ``Algorithm unrolling: Interpretable,
  efficient deep learning for signal and image processing,'' {\em IEEE Signal
  Process. Mag.}, vol.~38, no.~2, pp.~18--44, 2021.

\end{thebibliography}
\end{document}